\documentclass[twocolumn,tighten]{aastex62}

\usepackage{wasysym}

\newcommand{\sion}[2]{#1$\;$\textsc{#2}\relax}
\newcommand{\sub}[2]{\ifmmode #1_\mathrm{\scriptstyle #2} \else $#1_\mathrm{\scriptstyle #2}$\fi}
\shorttitle{\textit{HST}/COS Observations of Outflows in LBQS 1206+1052}
\shortauthors{Miller et al.}

\begin{document}

\title{Distance, Energy, and Variability of Quasar Outflows: Two \textit{HST}/COS epochs of LBQS 1206+1052}
\altaffiliation{Based on observations with the NASA/ESA \\\textit{Hubble Space Telescope} obtained at the Space Telescope \\Science Institute, which is operated by the Association \\of Universities for Research in Astronomy, Incorporated, \\under NASA contract NAS5-26555.}

\author[0000-0002-0730-2322]{Timothy R. Miller}
\affiliation{Department of Physics, Virginia Polytechnic Institute and State University, Blacksburg, VA 24061, USA}
\author{Nahum Arav}
\affiliation{Department of Physics, Virginia Polytechnic Institute and State University, Blacksburg, VA 24061, USA}
\author{Xinfeng Xu}
\affiliation{Department of Physics, Virginia Polytechnic Institute and State University, Blacksburg, VA 24061, USA}
\author{Gerard A. Kriss}
\affiliation{Space Telescope Science Institute, 3700 San Martin Drive, Baltimore, MD 21218, USA}
\author{Rachel J. Plesha}
\affiliation{Space Telescope Science Institute, 3700 San Martin Drive, Baltimore, MD 21218, USA}
\author{Chris Benn}
\affiliation{Isaac Newton Group, Apartado 321, E-38700 Santa Cruz del La Palma, Spain}
\author{Guilin Liu}
\affiliation{CAS Key Laboratory for Research in Galaxies and Cosmology, Department of Astronomy, University of Science and Technology of China, Hefei 230026, China}
\affiliation{School of Astronomy and Space Sciences, University of Science and Technology of China, Hefei 230026, China}



\begin{abstract}
We analyze new \textit{HST}/COS spectra for two quasar absorption outflows seen in the quasi-stellar object LBQS 1206+1052. These data cover, for the first time, absorption troughs from \sion{S}{iv}, \sion{Si}{ii}, and \sion{P}{v}. From the ratio of the \sion{S}{iv}* to \sion{S}{iv} column densities, we measure the electron number density of the higher-velocity ($-1400$~km~s$^{-1}$, v1400) outflow to be log(\sub{n}{e})~=~4.23$^{+0.09}_{-0.09}$~cm$^{-3}$ and constrain the lower-velocity ($-730$~km~s$^{-1}$, v700) outflow to log(\sub{n}{e})~$>$~5.3~cm$^{-3}$. The \sub{n}{e} associated with the higher-velocity outflow is an order of magnitude larger than reported in prior work. We find that the previous measurement was unreliable since it was based on density-sensitive absorption troughs that were likely saturated. Using photoionization models, we determine the best $\chi^2$-minimization fit for the ionization parameter and hydrogen column density of the higher-velocity outflow: log(\sub{U}{H})~=~$-1.73^{+0.21}_{-0.12}$ and log(\sub{N}{H})~=~$21.03^{+0.25}_{-0.15}$~cm$^{-2}$, respectively. We calculate from \sub{U}{H} and \sub{n}{e} a distance of 500$^{+100}_{-110}$~pc from the central source to the outflow. Using an SED attenuated by the v700 outflow yields a two-phase photoionization solution for the v1400 outflow, separated by a $\Delta U \approxeq 0.7$. Otherwise, the resultant distance, mass flux, and kinetic luminosity are similar to the unattenuated case. However, the attenuated analysis has significant uncertainties due to a lack of constraints on the v700 outflow in 2017.
\end{abstract}

\keywords{galaxies: active --- galaxies: kinematics and dynamics --- ISM: jets and outflows --- quasars: absorption lines --- quasars: general --- quasars: individual(LBQS 1206+1052)}

\section{Introduction}\label{sec:int}
A large fraction \citep[$\sim$20--40\% of all quasars;][]{hew03,dai08,kni08} of quasar spectra have blueshifted absorption troughs with respect to the rest frame of the quasar, indicative of outflowing material. These outflows are typically classified according to the widths of their absorption troughs. Widths greater than 2000~km~s$^{-1}$, between 500 and 2000~km~s$^{-1}$, and less than 500~km~s$^{-1}$ are labeled broad absorption lines (BALs), mini-broad absorption lines (mini-BALs), and narrow absorption lines (NALs), respectively \citep[e.g.][]{wey91,ald94,rei03,ves03,ham04,tru06}. These intrinsic outflows are prime candidates for producing active galactic nucleus (AGN) feedback processes: curtailing the growth of the host galaxy \citep[e.g.][]{cio09,hop09,sch15,cho17,pei17}, explaining the relationship between the masses of the central black hole and the host galaxy \citep[e.g.][]{sil98,bla04,boo09,hop09,dub14,ros15,vol16,ang17}, and chemical enrichment of the intergalactic and intracluster medium \citep[e.g.][]{sca04,kha08,tor10,bar11,tay15}. 
\\\\A crucial parameter needed to determine the impact of these outflows is their distance from the central source ($R$). These distances can be inferred from simultaneously determining the hydrogen number density (\sub{n}{H}) and ionization parameter (\sub{U}{H}) of the outflow. To date, $\sim$20 distances for quasar outflows have been determined using this method by our group \citep[e.g.][]{dek01,dek02,kor08,moe09,bau10,dun10,aok11,edm11,ara12,ara13,ara15,bor12,bor13,cha15a,cha15b,xu18} and others \citep[e.g.][]{ham01,gab05,fin14,luc14}. These distances are in the range of parsecs to tens of kiloparsecs, many orders of magnitude more distant than accretion disk wind models predict \citep[e.g.][]{mur95,pro00,pro04}.
\\\\AGN outflows also show variability on timescales of months to years \citep[e.g.][]{lun07,gib08,fil13,gri15,mcg17,mcg18}. This variability primarily manifests as changes in the depth of the trough with a few cases of velocity changes \citep[e.g.][]{ham97,vil01,gri16}. Two supported possibilities to account for these depth changes are a change in the ionizing photon rate incident on the outflow \citep[e.g.][]{ara15,gri15,wan15,wil15} or a change in the total hydrogen column density by material moving into the line of sight \citep[e.g.][]{ham08,hal11,viv12,cap13}. Recent work by \citet{he17} favors the former as the dominant cause for trough variability.
\\\\LBQS 1206+1052 was reported in the Large Bright Quasar Survey catalog by \citet{hew95}. Later, \citet{gib09} identified a \sion{Mg}{ii} ($\lambda \lambda$ 2796, 2803) trough in a Sloan Digital Sky Survey spectrum (data release 5), classifying it as a LoBAL quasar. Work by \citet{ji12} identified two outflows, v700 (centroid velocity $\approx$ $-730$ km s$^{-1}$) and v1400 (centroid velocity $\approx$ $-1400$ km s$^{-1}$), from \sion{Mg}{ii} and \sion{He}{i}* absorption troughs. These outflows cover a velocity width of $\sim$2000 km s$^{-1}$. \citet{cha15b} analyzed archival \textit{HST}/COS spectra from 2010 and determined the distance of the v1400 outflow from the central source to be 840 pc. There was concern, however, that the absorption troughs used to calculate the density could have been saturated, which can lead to an overestimation of $R$, prompting the follow-up observations in this work. 
\\\\\citet{sun17} presented a four-year (2012--2016) observing campaign of LBQS 1206+1052. Balmer absorption was identified for the v700 outflow, which yielded an estimate for the hydrogen number density in the range of 10$^9$--10$^{10}$ cm$^{-3}$ and a distance of $\sim$1 pc. Both the v1400 and v700 outflows showed absorption trough variability (only trough depth) consistent with the ionizing source changing. 
\\\\In this paper, we present in Section~\ref{sec:od} the new Hubble Space Telescope Cosmic Origins Spectrograph (\textit{HST}/COS) observations of LBQS 1206+1052, which cover a larger wavelength range compared to the 2010 observations. We also discuss in Section~\ref{sec:od} the spectral fitting for the continuum and emission lines. Section~\ref{sec:da} details the extraction of ionic column densities, photoionization modeling, and density calculations of the new observations as well as from the 2010 data. Our results and discussion on the physical properties, distances, energetics, and variability of the outflows are in Section~\ref{sec:rd}. A summary with conclusions is in Section~\ref{sec:sc}. We adopt an $h = 0.7$, $\Omega_m = 0.3$, and $\Omega_\Lambda = 0.7$ cosmology throughout this paper \cite[see, e.g.][]{ben14}.

\section{Observations, Data Reduction, and Spectral Fitting}
\label{sec:od}
\begin{figure*}
\includegraphics[scale=1.0]{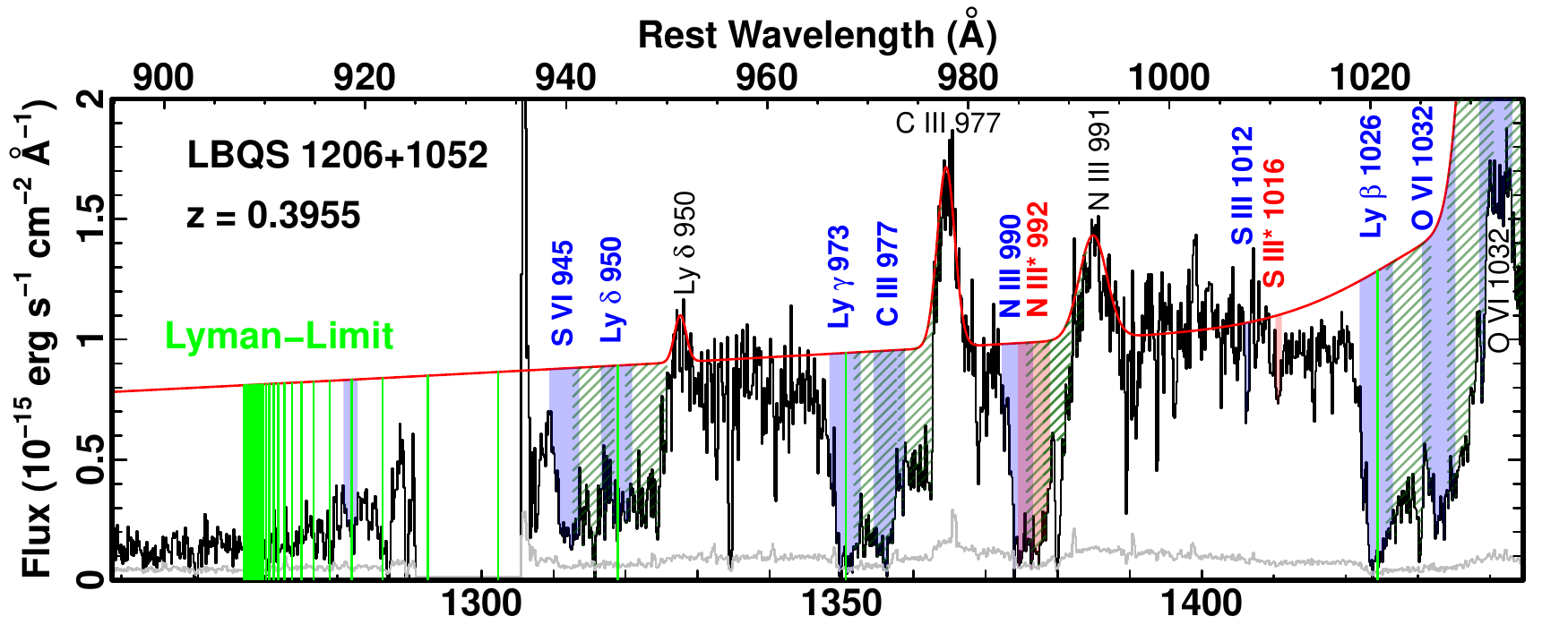}
\includegraphics[scale=1.0]{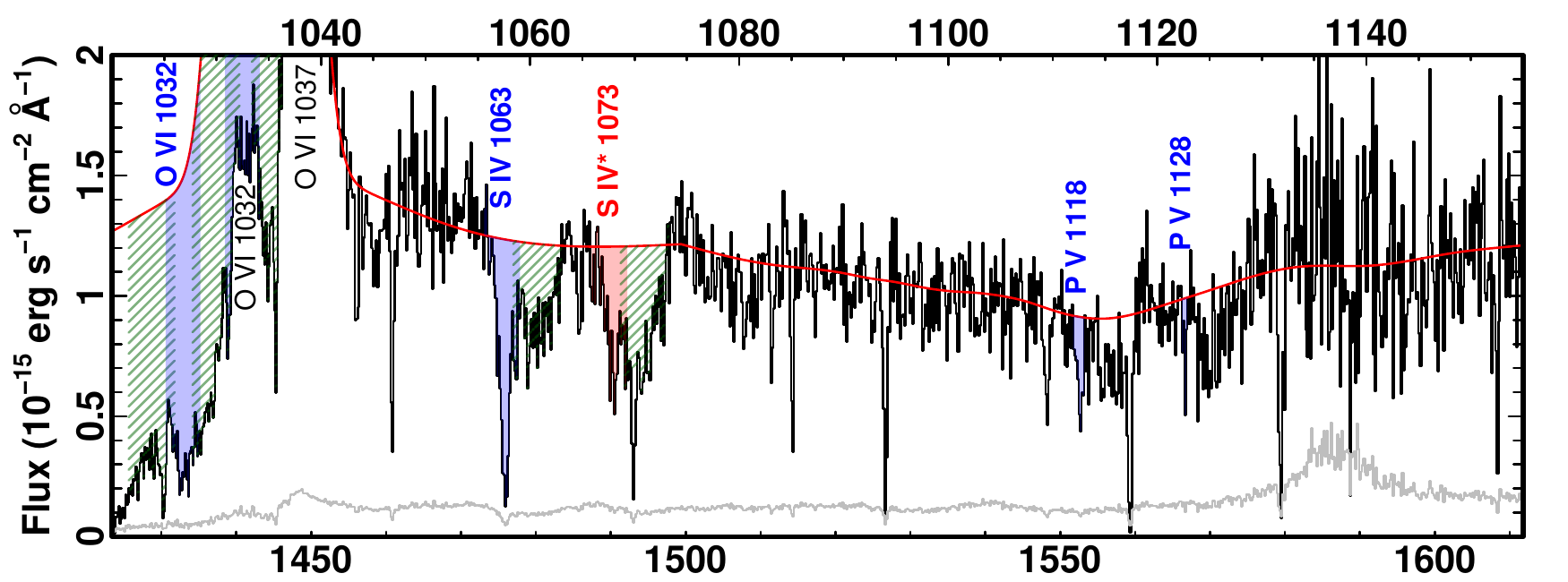}
\includegraphics[scale=1.0]{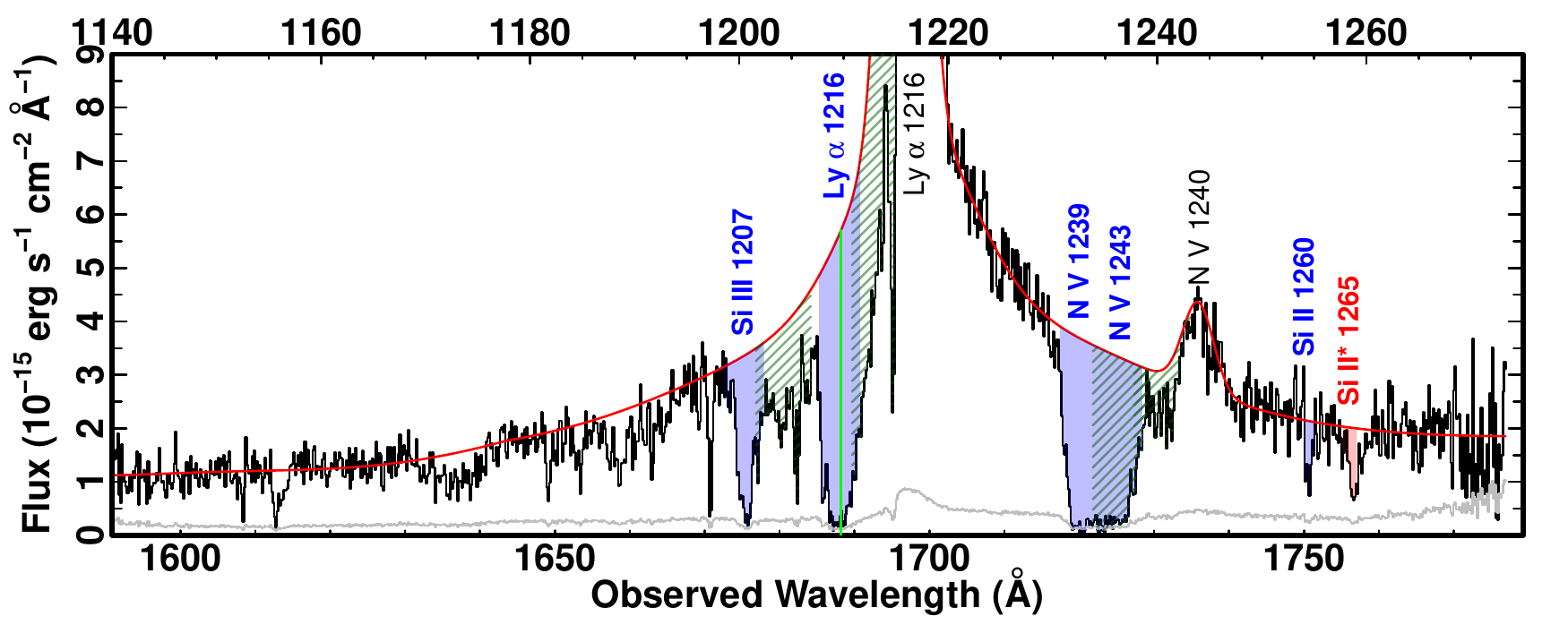}
\caption{\footnotesize{Portion of the \textit{HST}/COS 2017 spectrum (in black) showing the main absorption features. The blue and red shaded regions are the absorption trough transitions of the resonance and excited states, respectively, for the v1400 outflow. The slanted, dark green shaded regions show the absorption trough transitions for the v700 outflow. The red contour is the unabsorbed emission model and the green vertical lines indicate the Lyman series of hydrogen. The gray spectrum shows the errors, and emission lines are labeled in black.}}
\label{fig:spectrum}
\end{figure*}

LBQS 1206+1052 (J2000: R.A.~=~12~09~24.079, decl.~=~+10~36~12.06, $z$~=~0.3955) was first observed by \textit{HST}/COS in 2010 May (PID 11698) and again in 2017 July (PID 14777). The details of each observation are given in Table 1. The new observations cover a larger wavelength range, allowing for more density-sensitive diagnostics to be observed. For consistency, the 2010 data were reprocessed in the same way as the 2017 data, using up-to-date calibration files for each observation. The 2010 data were obtained at COS detector lifetime position 1 (LP1), for which the wavelength calibration is described in \citet{sts16} and the flat-field derivation in \citet{ely11}.  The 2017 data were obtained at LP3, and we used updated wavelength calibrations that are not yet publicly available to the science community. Flat fields appropriate to LP3 were derived at the new detector position using the methods described in \citet{ely11}. The data were reduced with a modified version of the STScI CALCOS v3.2.1 pipeline software. This version estimates errors for each pixel by the expression
\begin{equation}
err_i = \Bigg(\frac{N}{(s_it)^2}\Bigg)^{0.5}
\end{equation} where $err_i$ is the error for pixel $i$, $s_i$ is the sensitivity for pixel $i$, $N$ is the gross counts, and $t$ is the exposure time. Then, we binned each spectrum by 15 pixels and added the errors in quadrature. Finally, to render our Poisson-distributed errors as more similar to a Gaussian distribution, we added an additional count of flux to the errors, i.e.,
\begin{equation}
errG_i = err_i+(15s_it)^{-1}
\end{equation}where $errG_i$ is the Gaussian distributed error for binned position $i$, and $15$ is the bin size. 
\\\\The final one-dimensional spectrum for the 2017 data is shown in Figure~\ref{fig:spectrum}, while the spectrum for the 2010 data matches very well with Figure 11 of \citet{sun17}. There is a gap in the G130M detector around 1300 \AA. Absorption troughs from ions \sion{H}{i}, \sion{C}{iii}, \sion{N}{iii}, \sion{N}{v}, \sion{O}{vi}, \sion{Si}{ii}, \sion{Si}{iii}, \sion{P}{v}, \sion{S}{iii}, \sion{S}{iv}, and \sion{S}{vi}  are identified for both outflows in Figure~\ref{fig:spectrum} as well. Following the methodology of \citet{cha15b}, we fit the unabsorbed continuum emission with a power law and any emission lines with Gaussian profiles. The emission lines typically have both broad and narrow components originating from two distinct regions. The power law is of the form $F(\lambda)~=~F_{1100}(\lambda/1100)^\alpha$, where $F_{1100}=1.28\times10^{-15}$~erg~s$^{-1}$~cm$^{-2}$~\AA$^{-1}$ and $\alpha$~=~2.4. Since the absorption occurs on the blue side of the emission lines, the red side of each line was used to constrain the Gaussian fits. The centroid of each Gaussian was fixed at the rest-frame wavelength for each line, and lines originating from the same ions were modeled with the same FWHM. In the region where the power law did not match the spectrum (1500--1650 \AA, observed frame), a cubic spline was used to correct the fit. The solid red contour in Figure~\ref{fig:spectrum} shows the full, unabsorbed emission model adopted in this work.
\begin{deluxetable}{lccc}
\tablecaption{\textit{HST}/COS observations from 2010 and 2017 for LBQS 1206+1052.\label{tab:obs}}
\tablewidth{0pt}
\tabletypesize{\footnotesize}
\tablehead{
& \multicolumn{3}{c}{Date}
}
\startdata
 & 2010 May 8 & 2017 Jul 18 & 2017 Jul 18 \\
\tableline
\textit{HST}/COS Grating &  G130M & G130M & G160M \\
Exposure Time (s) & 4840 & 4320 & 4640 \\
Observed Range (\AA) & 1150--1445 & 1150--1445 & 1400--1780 \\
Rest-frame Range (\AA) & 825--1020 & 825--1020 & 1000--1275 \\
\enddata
\end{deluxetable}

\section{Data Analysis}
\label{sec:da}
\subsection{Ionic Column Density}
\label{sec:cd}
\begin{figure*}
\includegraphics[scale=0.33]{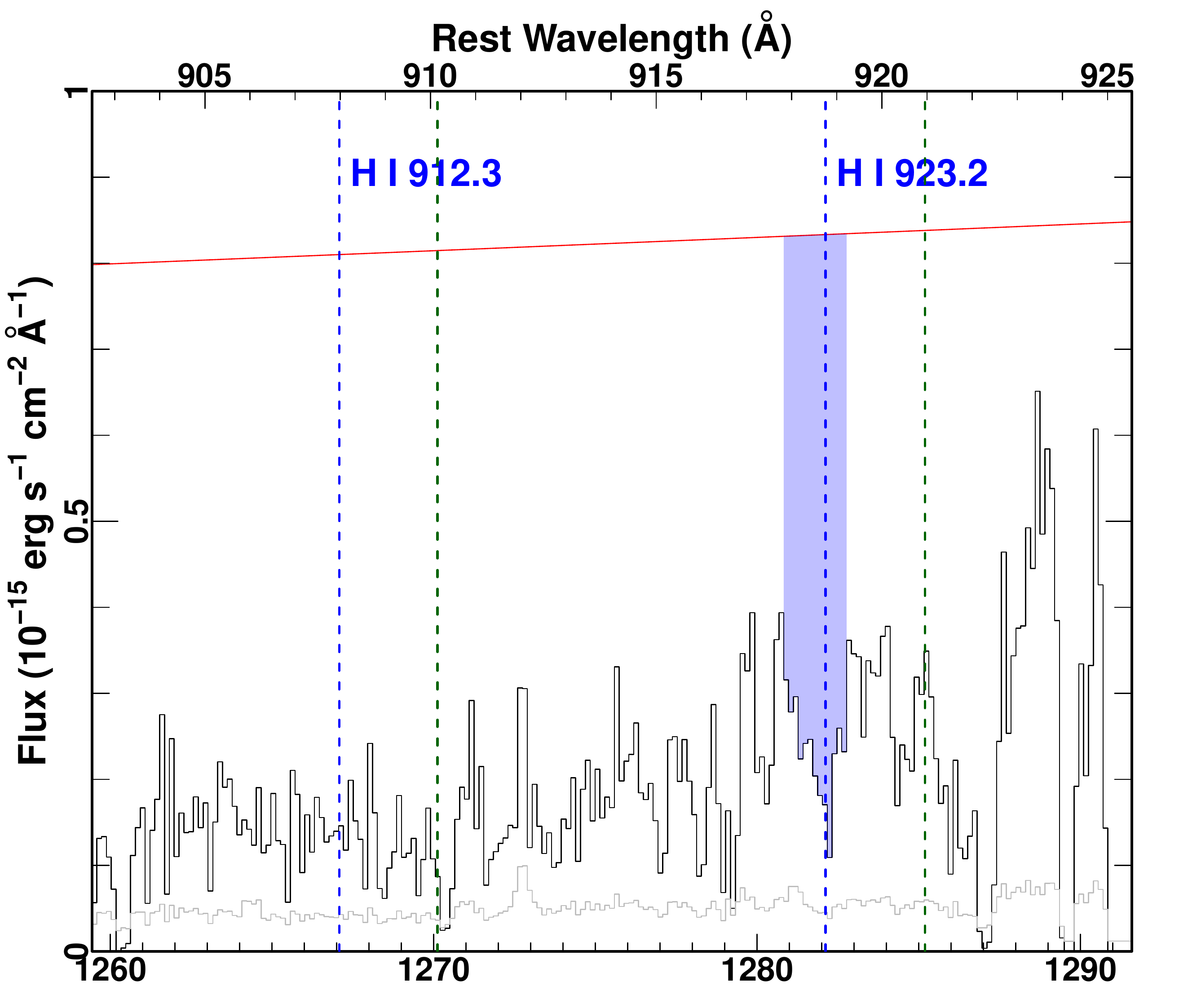}%
\includegraphics[scale=0.33]{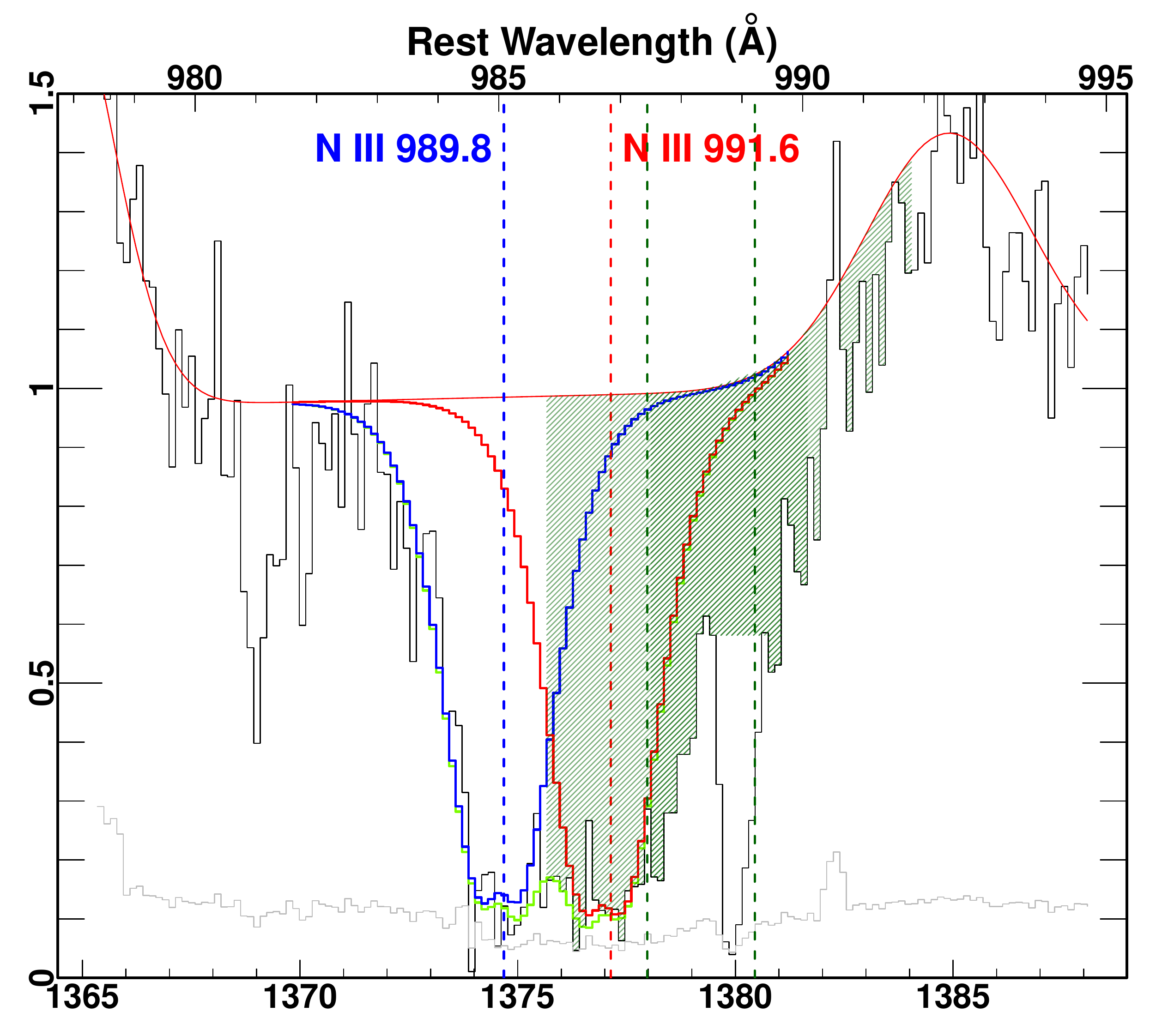}
\includegraphics[scale=0.33]{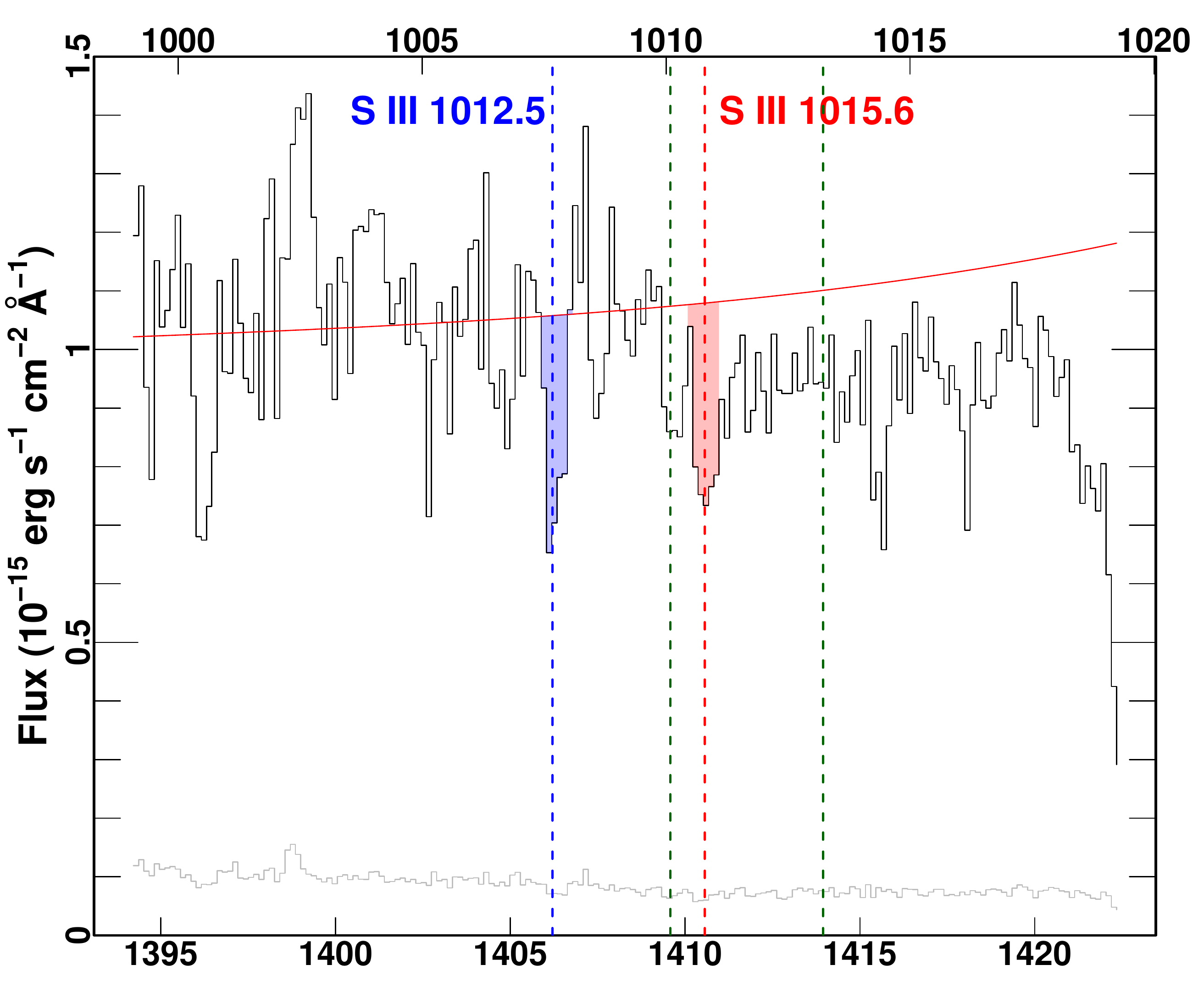}%
\includegraphics[scale=0.33]{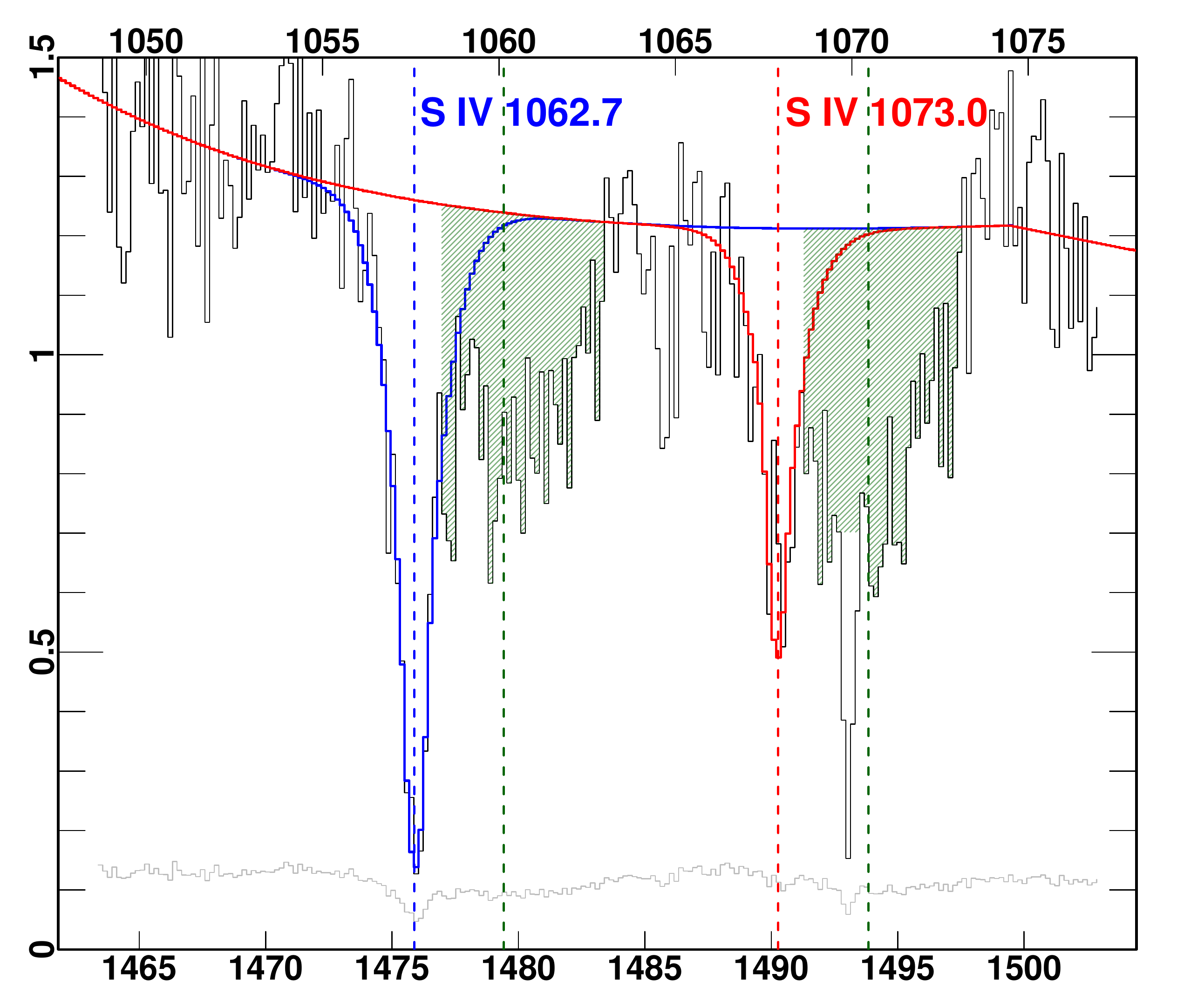}
\includegraphics[scale=0.33]{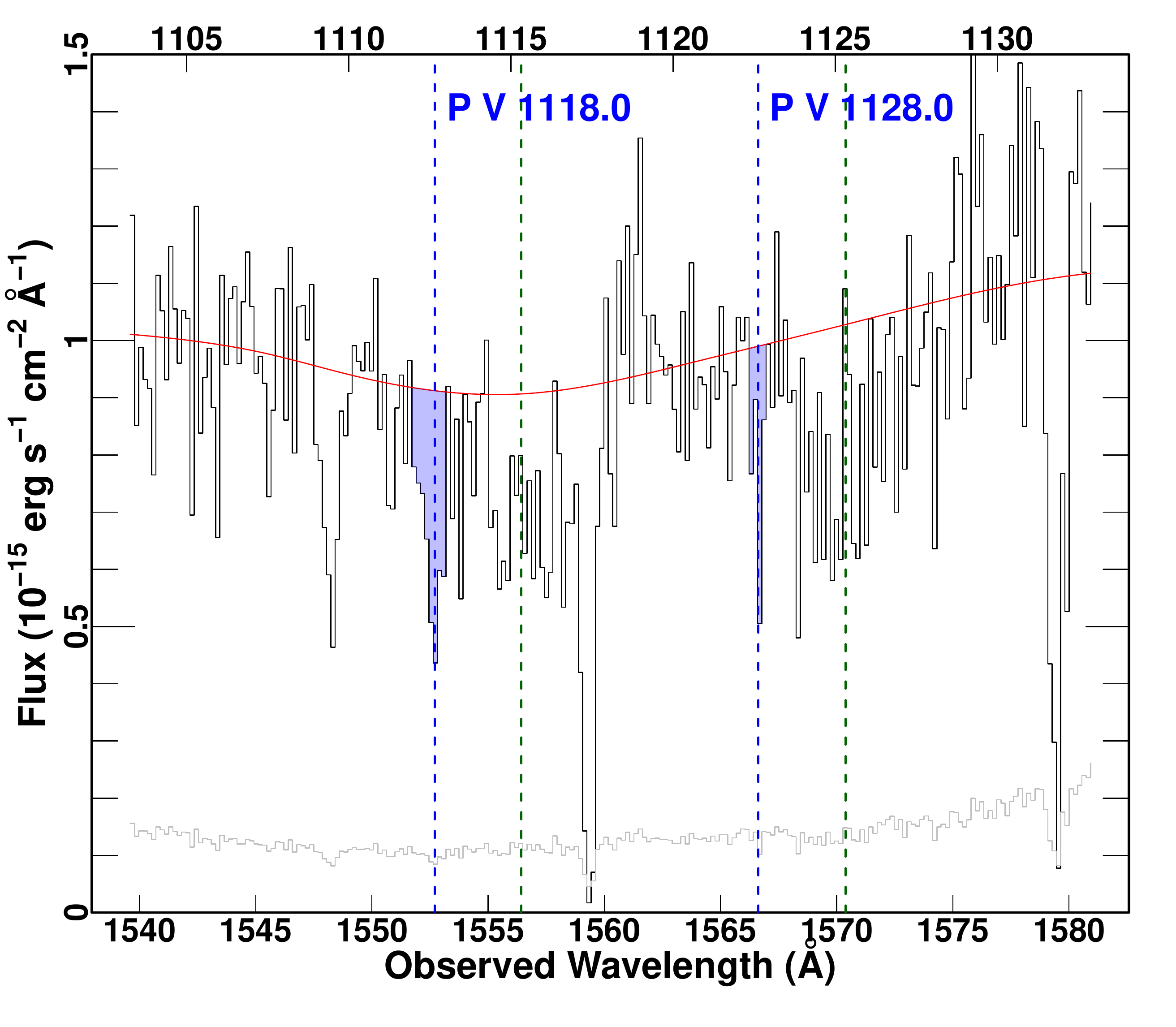}%
\includegraphics[scale=0.33]{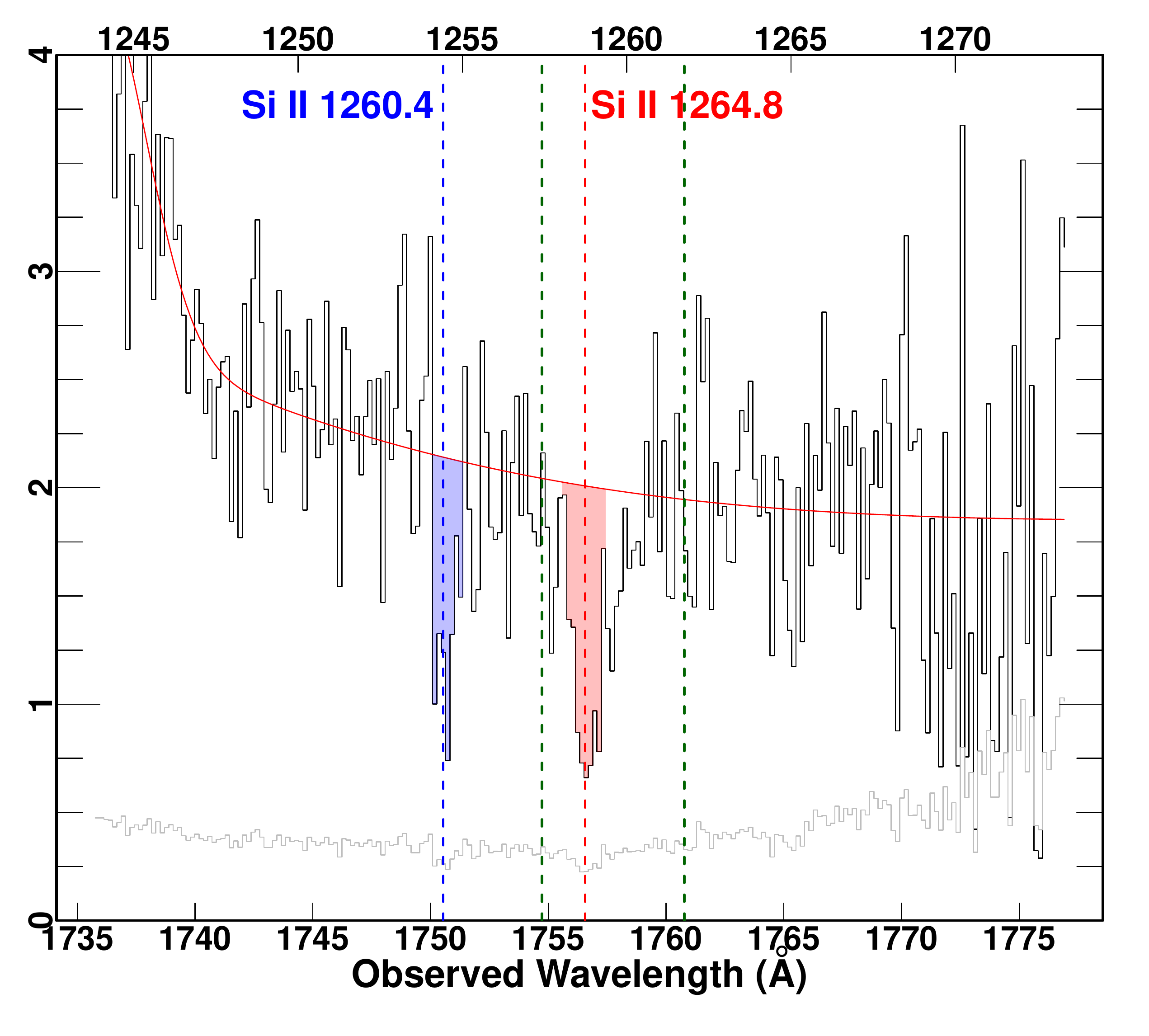}
\caption{\footnotesize{Zoomed-in portions of Figure~\ref{fig:spectrum} (same shaded regions and error spectrum). Overlaid on the \sion{N}{iii} and \sion{S}{iv} troughs are blue and red contours that are the components for the best-fit model of the v1400 outflow for each trough. For \sion{N}{iii}, the combination of the components is shown as a light green contour. The blue and red labels and similarly colored vertical, dashed lines identify the resonance and excited line transitions, respectively, of each ion. The vertical, green dashed lines show the v700 outflow counterparts to the v1400 outflow line transitions.}}
\label{fig:lines}
\end{figure*}

We measured the column density of a particular ion by using the apparent optical depth (AOD) method \citep{sav91}. This method assumes the outflow covers the source completely and homogeneously (i.e. the optical depth is the same across the source at each velocity). Therefore, for the optical depth, 
\begin{equation}
\tau(\lambda) \equiv -ln\Bigg(\frac{\sub{F}{obs}(\lambda)}{F_{0}(\lambda)}\Bigg)
\end{equation}
where $\sub{F}{obs}(\lambda)$ and $F_{0}(\lambda)$ are the observed flux and unabsorbed emission model flux, respectively. The ionic column density can then be solved from Equation (9) in \citet{sav91}:
\begin{equation}\label{eq:sav91}
\sub{N}{ion}=\frac{m_e c}{\pi e^2 f \lambda} \int \tau(v)dv
\end{equation} where $\sub{N}{ion}$ is the column density for an ionic transition, $m_e$ is the mass of the electron, $c$ is the speed of light, $e$ is the electric charge, $f$ is the oscillator strength for the ionic transition, $\lambda$ is the wavelength of the ionic transition, and $\tau(v)$ is the optical depth as a function of velocity. A variation of Equation (\ref{eq:sav91}) is used by \citet{cha15b} for blended troughs. AGN outflow troughs have been shown to suffer from non-black saturation, where the trough is saturated but the flux does not reduce to zero \citep[e.g.][]{ara08,bor12,bor13}. Typically, cases of non-black saturation are identified when measured optical depth ratios of two absorption lines from the same lower energy level deviate from the theoretical value determined by their oscillator strengths and wavelengths, i.e. $\frac{\tau_2}{\tau_1} \ne \frac{\lambda_2f_2}{\lambda_1f_1}$. In such cases, the AOD $\sub{N}{ion}$ measurement is treated as a lower limit. 
\\\\Figure~\ref{fig:lines} highlights the main troughs from the 2017 data, and the total ionic column density (excited plus resonance column densities) for each ion of the v1400 outflow is listed in Table~\ref{tab:col}. The predicted column densities from the best-fit model are also given (see Section~\ref{sec:photmod} and Figure~\ref{fig:sol}). The excited states for \sion{N}{iii}, \sion{S}{iii}, \sion{S}{iv}, and \sion{Si}{ii} have multiple transitions. These transitions have small separations so we combine each set of transitions into a single transition as labeled in Figure~\ref{fig:lines}. For the blended troughs of \sion{N}{iii} and \sion{S}{iv}, we employed Equation (1) from \citet{cha15b} to calculate the ionic column densities. The blue and red contours show the best component fits to each trough. The light green contour is the combination of the individual components for \sion{N}{iii}. As in Figure 1, the dark green, slanted lines show where the v700 absorption troughs should be located, contaminating each measurement. This contamination is minor in the case of \sion{S}{iv} since the fitting effectively removes it, but is potentially quite significant for \sion{N}{iii}. However, as we will discuss later in Section~\ref{sec:rd}, we believe the \sion{N}{iii} contamination is small and does not affect our results. 
\\\\For all other troughs in Figures~\ref{fig:spectrum}~\&~\ref{fig:lines}, we calculate the ionic column densities using Equation (\ref{eq:sav91}). The shaded blue and red regions show the integration range used for determining the ionic column densities of the resonance and excited state transitions, respectively. We estimate an upper limit for \sion{H}{i} from the Lyman limit near 912.3 \AA~in the same way as \citet{cha15b} with the expression $\tau=a_\nu \sub{N}{H}$, where $a_\nu$ is the photoionization cross section of \sion{H}{i}. As stated before, non-black saturation is a concern. For \sion{H}{i}, measurements closer to the Lyman limit are less saturated since the oscillator strength decreases. Therefore, the measurement at \sion{H}{i} 923.2 \AA~is the largest lower limit we reliably measure. Considering the very low abundance of phosphorus ($\approx$10$^{-3}$ times the abundance of carbon), we treat the \sion{P}{v} column density as a measurement. The column densities of \sion{O}{vi} and \sion{Si}{ii} are also taken as measurements for reasons discussed in Section~\ref{sec:rd}. All other measurements are taken as lower limits. We calculated the ionic column densities from the 2010 spectra in the same way. \sion{Ar}{iv}, like \sion{P}{v}, has a small abundance ($\approx$10$^{-2}$ times the abundance of carbon), and therefore is not likely saturated. In all measurements, the adopted values (see Table~\ref{tab:col}) assume errors of at least 20\% to account for systematic uncertainties in the unabsorbed emission model. These systematic uncertainties arise from features like those seen in Figure~\ref{fig:spectrum} around the \sion{S}{iv} absorption troughs, where shifting the continuum to better match those features results in total \sion{S}{iv} column density errors near 20\%.

\begin{deluxetable}{cccc}
\tablecaption{Total Ionic Column Densities in Component v1400\label{tab:col}}
\tablewidth{0pt}
\tabletypesize{\footnotesize}
\tablehead{
\colhead{Ion} & \colhead{Measured} & \colhead{Adopted} & \colhead{Best Model}\\
 & \colhead{($10^{13} $cm$^{-2}$)} & \colhead{($10^{13} $cm$^{-2}$)} & \colhead{($10^{13} $cm$^{-2}$)}
}
\startdata
\multicolumn{4}{c}{2017 Data}\\
\tableline
\sion{H}{i} Upper Limit & 28,000$^{+2800}_{-2800}$ & 28,000$^{+5600}_{}$ & 33,000\\
\sion{H}{i} Lower Limit & 12,000$^{+690}_{-530}$ & 12,000$^{}_{-2400}$ & 33,000\\
\sion{N}{iii} & 710$^{+60}_{-60}$ & 710$^{}_{-140}$ & 2550\\
\sion{N}{v} & 510$^{+55}_{-20}$ & 510$^{}_{-100}$ & 740\\
\sion{C}{iii} & 82$^{+5}_{-3}$ & 82$^{}_{-16}$ & 10,800\\
\sion{O}{iii} & 1400$^{+310}_{-160}$ & 1400$^{}_{-280}$ & 31,500\\
\sion{O}{vi} & 430$^{+8}_{-7}$ & 430$^{+430}_{-86}$ & 1070\\
\sion{Si}{ii} & 9.0$^{+1}_{-0.8}$ & 9.0$^{+1.8}_{-1.8}$ & 8.1\\
\sion{Si}{iii} & 18$^{+1.0}_{-0.7}$ & 18$^{}_{-3.6}$ & 230\\
\sion{P}{v} & 8.2$^{+1.4}_{-1.1}$ & 8.2$^{+1.6}_{-1.6}$ & 6\\
\sion{S}{iii} & 97$^{+9}_{-8}$ & 97$^{}_{-19}$ & 200\\
\sion{S}{iv} & 630$^{+60}_{-60}$ & 630$^{}_{-130}$ & 610\\
\sion{S}{vi} & 230$^{+11}_{-8}$ & 230$^{}_{-47}$ & 230\\
\tableline
\multicolumn{4}{c}{2010 Data}\\
\tableline
\sion{H}{i} Upper Limit & 25,000$^{+830}_{-790}$ & 25,000$^{+5000}_{}$ & 26,500\\
\sion{H}{i} Lower Limit & 16,000$^{+1100}_{-860}$ & 16,000$^{}_{-3200}$ & 26,500\\
\sion{N}{iii} & 690$^{+31}_{-31}$ & 690$^{}_{-140}$ & 2080\\
\sion{C}{iii} & 93$^{+6}_{-3}$ & 93$^{}_{-19}$ & 8810\\
\sion{O}{iii} & 1700$^{+320}_{-130}$ & 1700$^{}_{-340}$ & 24,700\\
\sion{O}{vi} & 410$^{+15}_{-11}$ & 410$^{+410}_{-82}$ & 940\\
\sion{S}{iii} & 89$^{+16}_{-13}$ & 89$^{}_{-18}$ & 160\\
\sion{S}{vi} & 210$^{+9}_{-7}$ & 210$^{}_{-42}$ & 200\\
\sion{Ar}{iv} & 160$^{+94}_{-42}$ & 160$^{+94}_{-42}$ & 120
\enddata
\tablecomments{Total ionic column densities (excited plus resonance, where applicable, of the v1400 outflow) for each observation with the measured and adopted errors. The predicted column densities from the best-fit Cloudy model are in the last column. Since the flux decreased from 2010 to 2017, \sion{Ar}{iv} was not reliably measured for the 2017 data.}
\end{deluxetable}

\subsection{Photoionization Modeling}
\label{sec:photmod}
\begin{figure}
\includegraphics[scale=0.305]{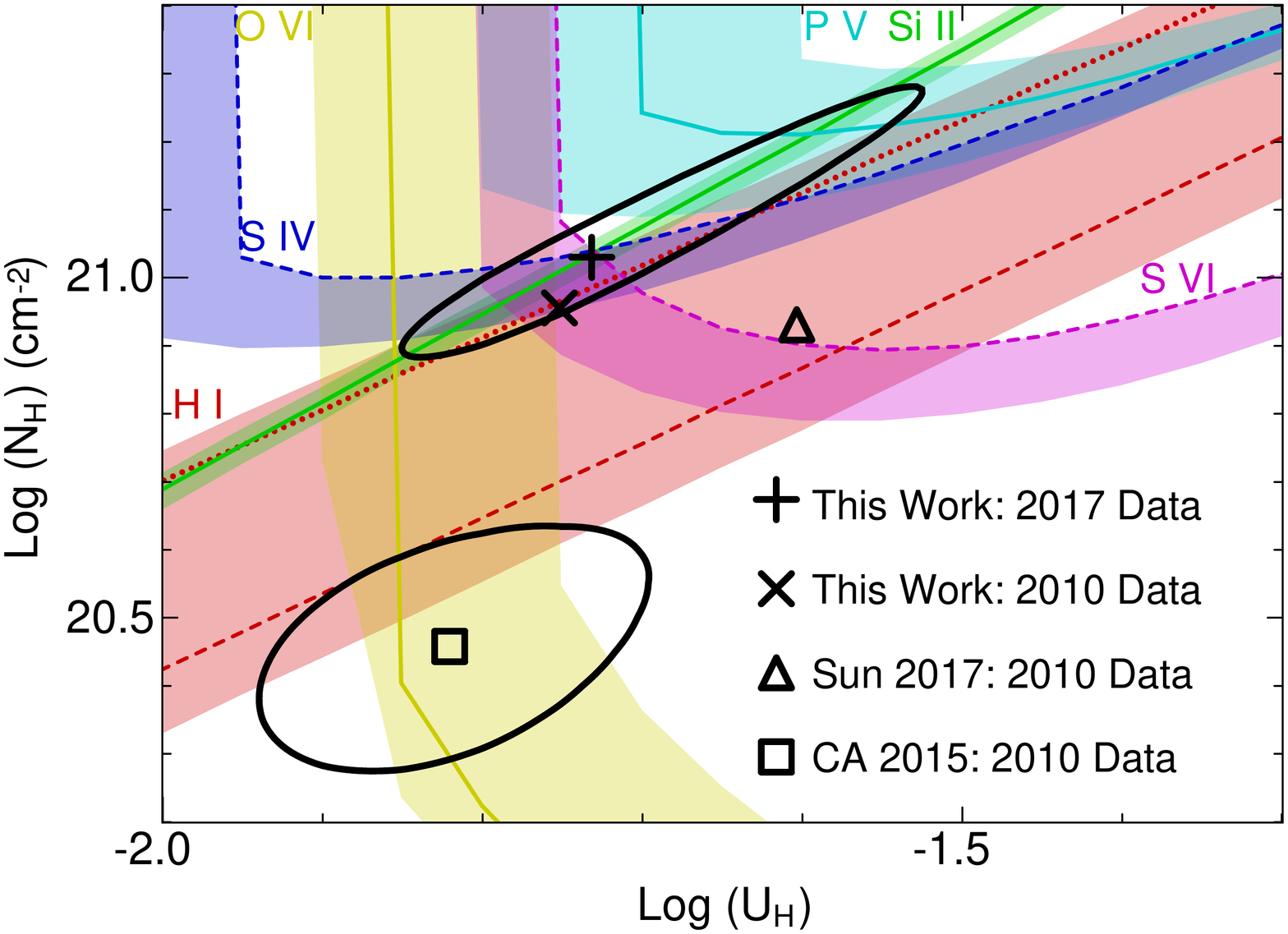}\\
\includegraphics[scale=0.305]{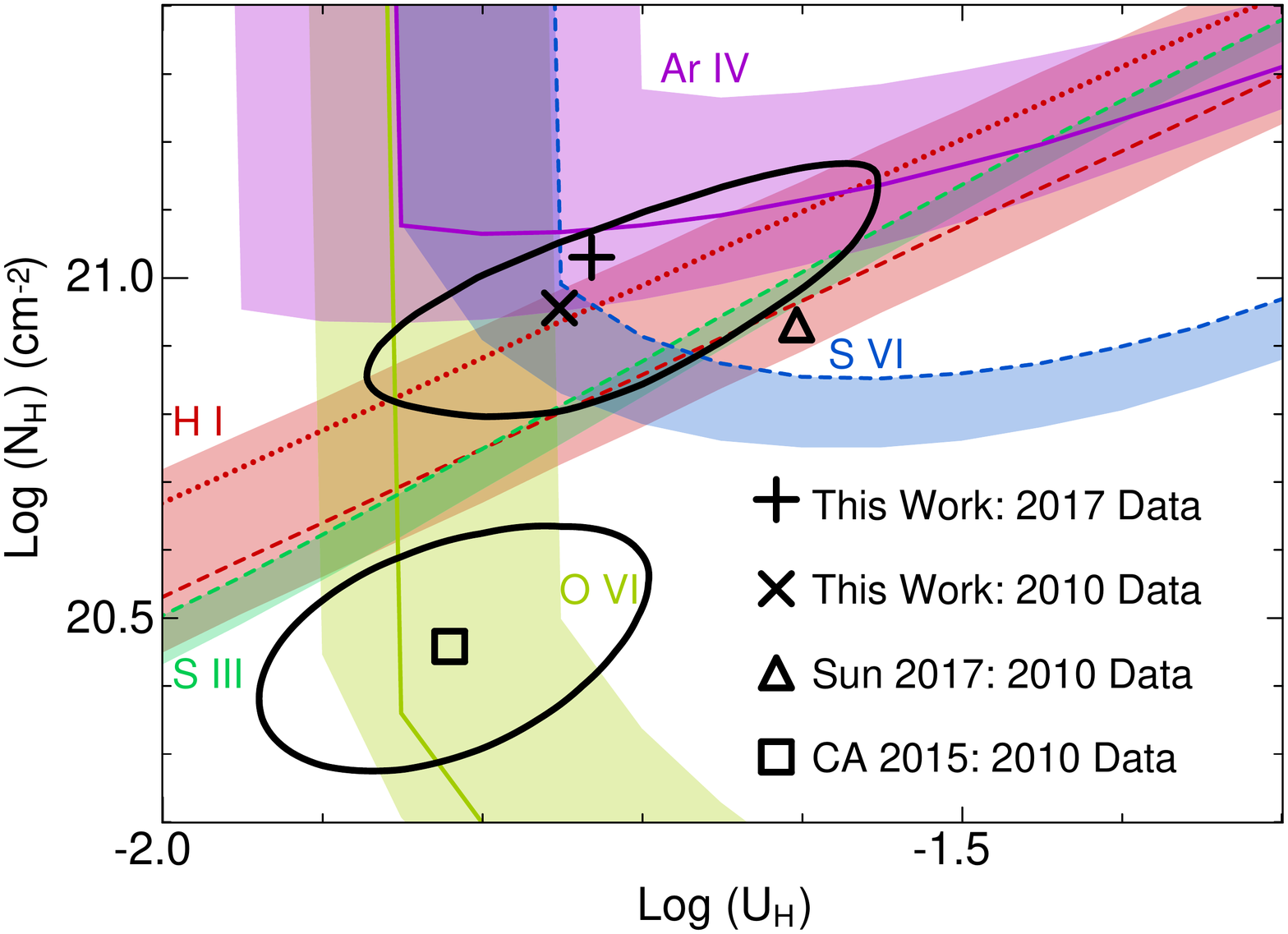}
\caption{\footnotesize{\textit{Top:} photoionization solution for the v1400 outflow based on the 2017 data. \textit{Bottom:} photoionization solution for the v1400 outflow based on the reprocessed 2010 data. \textit{For both panels:} the colored contours show the model parameters that are consistent with the observed value. Solid, dotted, and dashed contours represent ionic column densities taken as measurements, upper limits, and lower limits, respectively. The shaded bands are the 1$\sigma$ uncertainties for each contour (see Table~\ref{tab:col}). The solutions from \citet{cha15b}, CA 2015, and \citet{sun17}, Sun 2017, are also shown. The plus, cross, triangle, and square symbols are the best $\chi^2$-minimization solutions and the black ellipses (not available for Sun 2017) encircling them are their 1$\sigma$ uncertainties.}}
\label{fig:sol}
\end{figure}

The ionization structure of the outflow determines the ionic column densities we measured. Therefore, by using the ionic column densities in conjunction with a grid of Cloudy \citep[][version c17.00]{fer17} photoionization models, a solution for the hydrogen column density (\sub{N}{H}) and ionization parameter can be found. For each model, we assumed solar metallicity \citep{gre10}, the UV-soft SED \citep{dun10}, and a stopping criterion that the proton to hydrogen density ratio equals 0.01 (ensuring a wide range of \sub{N}{H}). The UV-soft SED was chosen since the overall UV--optical spectrum shown in Figure 13 of \citet{sun17} resembles AGN SEDs illustrated in \citet{sha05,sha11}, which are similar to the UV-soft SED. In this section, we do not take into account the effects the v700 outflow has on the SED before reaching the v1400 outflow \citep[see][their ``shading effect"]{sun17}. This is because the properties of the v700 outflow have not been determined robustly since the Balmer, \sion{He}{i*}, and \sion{Mg}{ii} absorption troughs were not covered in the 2017 epoch. However, in Section \ref{sec:tp}, we show that when the v700 properties from \citet{sun17} are assumed, the distance, mass flux, and kinetic luminosity for the v1400 outflow are only mildly affected by this ``shading effect."
\\\\For a particular pair of \sub{N}{H} and \sub{U}{H}, ionic column densities from the model are compared to the measured counterparts. In Figure~\ref{fig:sol}, the colored contours for individual ions show where the model predicted ionic column densities are consistent ($<$ 1$\sigma$) with the observed values. The colored contours with solid, dotted, or dashed lines show the ionic column densities treated as measurements, upper limits, or lower limits, respectively. The best-fit solution is determined through $\chi^2$-minimization of the model predicted ionic column densities compared to the measured, ionic column densities (all values in Table~\ref{tab:col}). The solutions and corresponding 1$\sigma$ uncertainties are the black symbols and ellipses, respectively. 

\subsection{Electron Number Density}
\label{sec:ed}
\begin{figure}
\includegraphics[scale=0.34]{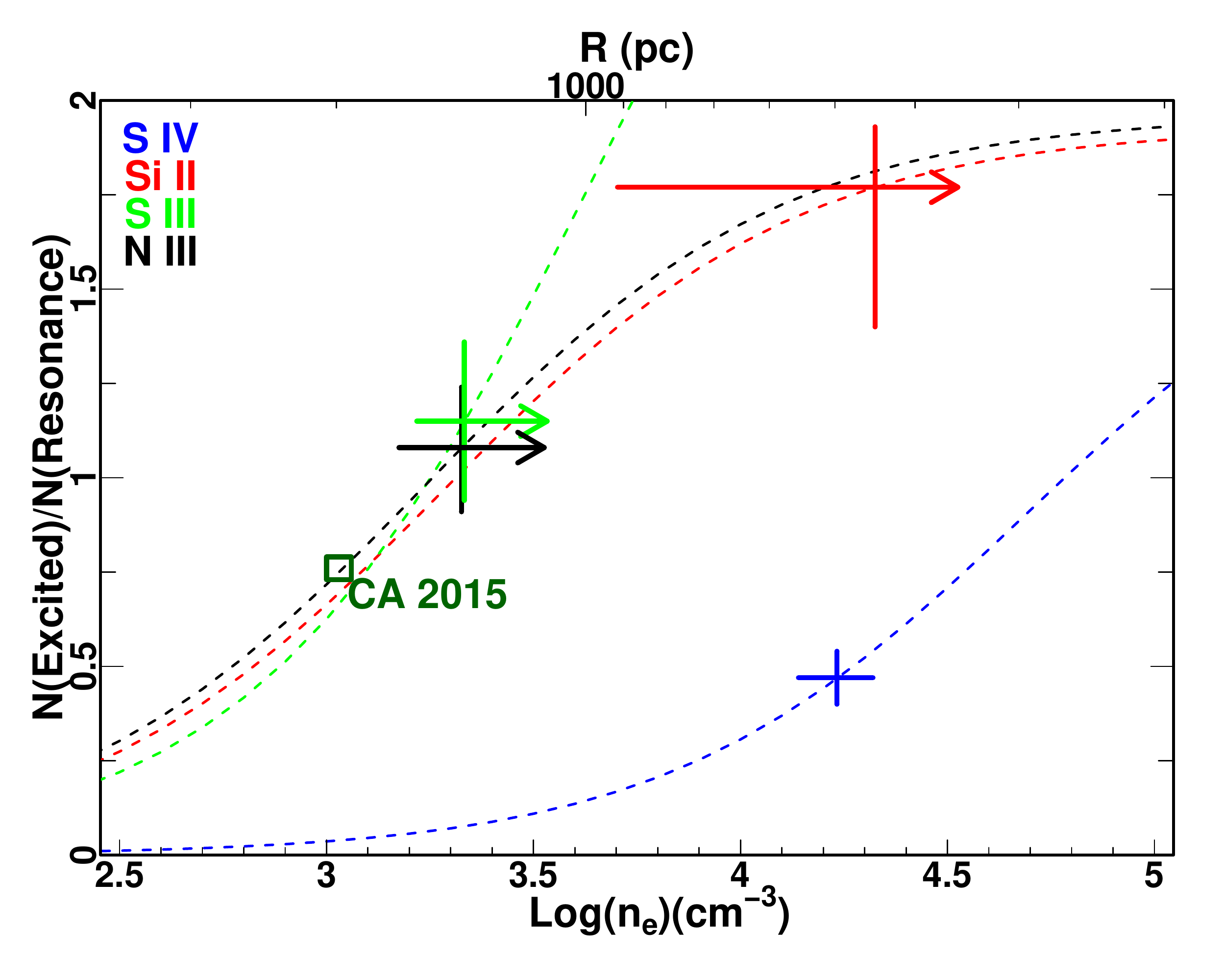}\\
\includegraphics[scale=0.34]{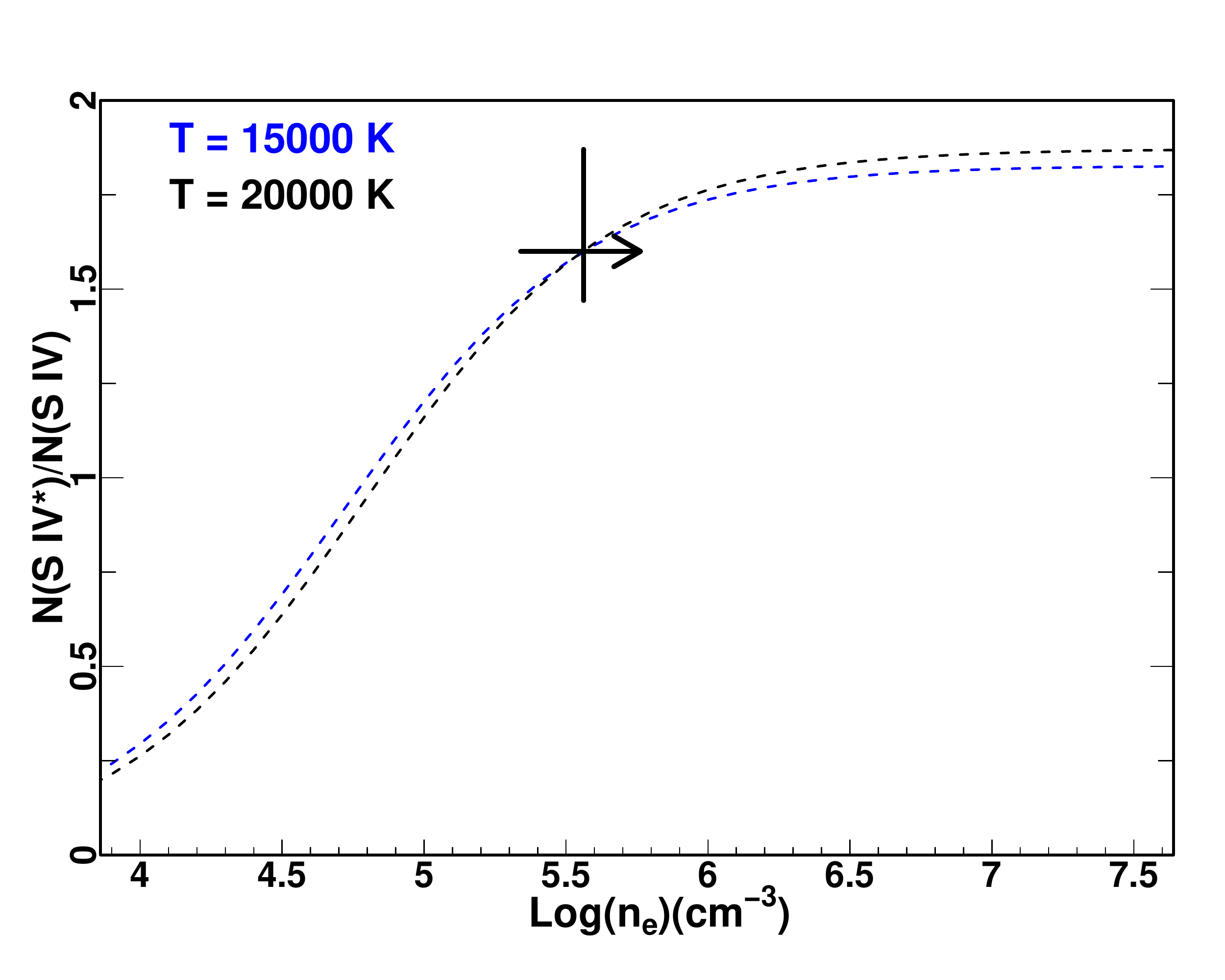}
\caption{\footnotesize{\textit{Top:} electron number density, \sub{n}{e}, and distance from the central source, $R$ (Equation (\ref{eq:R})), of the v1400 outflow based on the 2017 data column density ratios of \sion{S}{iv}, \sion{Si}{ii}, \sion{S}{iii}, and \sion{N}{iii}. The dashed contours are the expected ratios for each ion at a given density for a temperature of 13,500 K, the average temperature from the photoionization solution. The arrows indicate lower limits. The solution for \sion{N}{iii} from \cite{cha15b} is also shown. \textit{Bottom:} the lower limit to the electron number density for the v700 outflow based on the \sion{S}{iv} ratio. The dashed contours show the expected ratio for a given density for two temperatures, encompassing the average temperatures based on the photoionization solutions of \citet{sun17}. This lower limit is consistent with the \citet{sun17} range: $\sub{n}{e} = 10^9$--$10^{10}$ cm$^{-3}$}}
\label{fig:dens}
\end{figure}

The energy level for the excited states of \sion{S}{iv}, \sion{S}{iii}, \sion{Si}{ii}, and \sion{N}{iii} are populated by collisional excitations with free electrons. Therefore, the relative populations between the excited and resonance states depend on the electron number density, \sub{n}{e}. There is a small temperature dependence, but the range in temperatures allowed by the best-fit photoionization solution is less than 1000 K (introducing only a few percent error in \sub{n}{e}). Following the methodology of previous works \citep[e.g.][]{kor08,bor12b,ara13,ara18,cha15b}, we used the CHIANTI 8.0.7 database \citep{der97,lan13} to calculate the predicted population ratios of the excited to resonance states for each ion. This ratio is equal to the ratio of the column densities of the excited to resonance states. The $\sub{N}{ion}$ measured values are given in Table~\ref{tab:ratios}.
\\\\In the top panel of Figure~\ref{fig:dens}, the colored, dashed contours show the expected column density ratio as a function of electron number density for each ion at the temperature of the 2017 photoionization solution, 13,200 K. Overlaid on the contours are the measured column density ratios and uncertainties of each ion from the 2017 data (Table~\ref{tab:ratios}). As can be seen, the column density ratios for \sion{N}{iii} and \sion{S}{iii} are consistent with 1, indicative of saturation \citep{bor13, ara18}. This means the value of the true ratio could be greater than or less than 1, depending on if the excited or resonance trough is more heavily saturated \citep{ara18}. The model predicted total column densities for \sion{N}{iii} and \sion{S}{iii} in Table~\ref{tab:col} supports the saturation claim since they are 2--3 times larger than each respective measured value. Also, the upper error for the \sion{Si}{ii} column density ratio is large enough that the ratio is consistent with the asymptotic maximum value. Therefore, the only reliable measurement of the electron number density is from the \sion{S}{iv} column density ratio, which is firmly below 1. We calculate a value for the electron number density of log(\sub{n}{e}) = 4.23$^{+0.09}_{-0.09}$~cm$^{-3}$ based on the \sion{S}{iv} column density ratio and treat the measurements from the other column density ratios as lower limits.

\begin{deluxetable}{lccc}
\tabletypesize{\footnotesize}
\tablecaption{Excited and Resonance Ionic Column Densities\label{tab:ratios}}
\tablewidth{0pt}
\tablehead{
\colhead{Column Density} & \colhead{v1400} & \colhead{v700} & \colhead{v1400}\\
\colhead{(10$^{13}$ cm$^{-2}$)} & \colhead{2017 Data} & \colhead{2017 Data} & \colhead{2010 Data}
}
\startdata
$N$(\sion{S}{iv}*) & 200$^{+21}_{-21}$ & 400$^{+24}_{-21}$ & \nodata\\
$N$(\sion{S}{iv}) & 430$^{+50}_{-50}$ & 250$^{+17}_{-15}$ & \nodata\\
$N$(\sion{Si}{ii}*) & 5.8$^{+0.9}_{-0.7}$ & \nodata & \nodata \\
$N$(\sion{Si}{ii}) & 3.3$^{+0.5}_{-0.5}$ & \nodata & \nodata \\
$N$(\sion{S}{iii}*) & 52$^{+6}_{-5}$ & \nodata & 52$^{+11}_{-9}$\\
$N$(\sion{S}{iii}) & 45$^{+7}_{-6}$ & \nodata & 55$^{+11}_{-9}$\\
$N$(\sion{N}{iii}*) & 370$^{+30}_{-30}$ & \nodata &  360$^{+24}_{-24}$\\
$N$(\sion{N}{iii}) & 340$^{+50}_{-50}$ & \nodata & 330$^{+20}_{-20}$\\
\enddata
\tablecomments{Excited and resonance ionic column densities of the density-sensitive ions for the v1400 and v700 outflows in LBQS~1206+1052.}
\end{deluxetable}

\section{Results and Discussion}
\label{sec:rd}
\subsection{v1400 Outflow Properties, Distance, and Energetics}
\label{v1400opde}
Using the methodology of \citet{bor13}, we determine the distance from the central source to the v1400 outflow from the ionization parameter: 
\begin{equation}
\label{eq:R}
\sub{U}{H} = \frac{\sub{Q}{H}}{4\pi R^2\sub{n}{H} c}
\end{equation}
where $R$ is the distance from the central source, \sub{n}{H}~is the hydrogen number density ($\sub{n}{e} = 1.2\sub{n}{H}$ for highly ionized plasma), $c$ is the speed of light, and $\sub{Q}{H}$ is the ionizing hydrogen photon rate. We calculate $\sub{Q}{H}$ by integrating the UV-soft SED, normalized to each epoch, for energies above 1 Ryd. \citet{sun17} showed that the UV portion of the spectrum is likely heavily dust extincted. We concur with this assessment (see Section \ref{intdustext}). Therefore, we chose to scale the UV-soft SED for the 2010 data to the same normalization as \citet{sun17}, $L_{3000}$~=~4.7$\times$10$^{45}$ erg s$^{-1}$. For the 2017 data, we scaled the normalization by the amount the continuum level decreased between the 2010 and 2017 epochs ($\sim$20\%). This yields $\sub{Q}{H}=2.45\times 10^{56}$~s$^{-1}$ and $\sub{Q}{H}=3.09\times 10^{56}$~s$^{-1}$ for the 2017 data and 2010 data, respectively. Following the assumptions of \citet{bor12}, the mass flow rate and kinetic luminosity are given by the following expressions:
\begin{equation}
\label{eq:M}
\dot{M}\simeq 4\pi \Omega R N_H \mu m_p v 
\end{equation}
\begin{equation}
\label{eq:E}
\dot{\sub{E}{K}}\simeq \frac{1}{2} \dot{M} v^2
\end{equation}
where the global covering factor $\Omega = 0.08$ \citep[fraction of outflows with \sion{S}{iv}/\sion{S}{iv}* troughs;][]{bor13}, $R$ is the distance from the central source, $\mu = 1.4$ is the mean atomic mass per proton, \sub{N}{H} is the hydrogen column density, $m_p$ is the proton mass, and $v$ is the outflow velocity.
\\\\Table~\ref{tab:comp} contains the physical properties, energetics, and distances for the v1400 outflow as determined by this work, \citet{cha15b}, and \citet{sun17}. Since the column density ratios and total column densities (see Tables~\ref{tab:col}~and~\ref{tab:ratios}) of \sion{N}{iii} and \sion{S}{iii} indicate saturation, and they were unaware of the dust extinction, the results from \citet{cha15b} differ significantly from ours. However, their main conclusion that this outflow is not a major contributor to AGN feedback still holds. The main difference between our work and \citet{sun17}, besides the number density, is the choice of SED. \citet{sun17} employed the MF87 SED \citep{mat87}, which is harder in the UV compared to the UV-soft SED. 
Choosing the MF87 SED results in a $\sub{Q}{H}$ over two times larger than ours, which when combined with the different electron densities, gives the much larger distance they calculate.
\begin{deluxetable}{lcccc}
\tabletypesize{\footnotesize}
\tablecaption{Comparison: Physical Properties, Distances, and Energetics of the v1400 Component\label{tab:comp}}
\tablewidth{0pt}
\tablehead{
\colhead{Analysis} & \colhead{TW} & \colhead{TW} & \colhead{CA} & \colhead{Sun} \\
\tableline
\colhead{Data} & \colhead{2017} & \colhead{2010} & \colhead{2010} & \colhead{2010}
}
\startdata
log$_{}$(\sub{N}{H}) & 21.03$^{+0.25}_{-0.15}$ & 20.96$^{+0.21}_{-0.16}$ & 20.46$^{+0.17}_{-0.17}$ & 20.93 \\
\ [cm$^{-2}$]\\
\tableline
log$_{}$(\sub{U}{H}) & -1.73$^{+0.21}_{-0.12}$ & -1.75$^{+0.20}_{-0.12}$ & -1.82$^{+0.12}_{-0.12}$ & -1.60 \\
\ [dex]\\
\tableline
log(\sub{n}{e}) & 4.23$^{+0.09}_{-0.09}$ & \nodata & 3.03$^{+0.06}_{-0.06}$ & 3.08 \\
\ [cm$^{-3}$]\\
\tableline
Distance & 500$^{+100}_{-110}$ & \tablenotemark{a}590$^{+120}_{-130}$ & 840$^{+60}_{-60}$ & 2800 \\
\ [pc]\\
\tableline
$\dot{M}$ & 8.9$^{+7.2}_{-3.1}$ & \tablenotemark{a}8.8$^{+5.8}_{-3.2}$ & 9.0$^{+3.0}_{-3.0}$ & 40 \\
\ [$M_{\astrosun}$yr$^{-1}$]\\
\tableline
log($\dot{\sub{E}{K}}$) & 42.8$^{+0.26}_{-0.19}$ & \tablenotemark{a}42.8$^{+0.22}_{-0.20}$ & 42.8$^{+0.15}_{-0.15}$ & 43.4\\
\ [erg s$^{-1}$]\\
\enddata
\tablecomments{Comparison between this work, TW, \citet{cha15b}, CA, and \citet{sun17}, Sun, of the physical properties, distances, and energetics of the v1400 outflow in LBQS 1206+1052.}
\tablenotetext{a}{Assumes the 2017 data \sub{n}{e} measurement.}
\end{deluxetable}
\subsection{Internal Dust Extinction}
\label{intdustext}
As stated in Section \ref{v1400opde}, \citet{sun17} provided evidence that heavy dust extinction is present in the 2010 data. They noted that the emission line profiles of \sion{C}{iii} $\lambda$977, \sion{N}{iii} $\lambda$992, and \sion{O}{vi} $\lambda$1031 had components too wide to be from the narrow emission line region (typical FWHM $\sim$500 km s$^{-1}$) and too narrow to be from the broad emission line region. This type of situation can occur when the broad emission lines are suppressed by dust extinction from the torus and intermediate-width emission lines from the inner face of the torus begin to dominate the observed emission line profiles \cite[see, e.g.,][]{li15}. 
\\\\Similar intermediate-width emission lines are still seen in the 2017 data. The fits to the emission lines of Ly $\lambda$1216, \sion{O}{vi} $\lambda$1032, and \sion{O}{vi} $\lambda$1037 in the unabsorbed emission model of Figure~\ref{fig:spectrum} comprise one weak but broad ($\sim$20\% of max height, FWHM $\sim$6000 km s$^{-1}$ with errors $<$20\%) and one strong yet still wide ($\sim$80\% of max height, FWHM $\sim$1000 km s$^{-1}$ with errors $<$15\%) Gaussian components. The former component is wide enough to be originating from the broad emission line region, but the latter component is too wide to be coming from the narrow emission line region. The fits to the \sion{C}{iii} $\lambda$977 and \sion{N}{iii} $\lambda$992 emission lines also only require a single Gaussian component with an FWHM $\sim$1000 km s$^{-1}$ (with errors $<$20\%). The persistence of these intermediate-width emission lines and disappearance of the broad-line components blueward of \sion{O}{vi} $\lambda$1032 lead us to believe there is heavy dust extinction in the UV spectrum.
\subsection{Photoionization Variability}
In Section~\ref{sec:int}, we mentioned two possible causes of variability for outflow absorption troughs seen in quasar spectra, a change in ionizing flux or material moving into the line of sight, and that \citet{sun17} supported the former. As can be seen in Figure~\ref{fig:sol}, the 2010 and 2017 photoionization solutions are compatible with either cause. However, given the arguments of \citet{sun17} that the absorption trough depths both increased and decreased over time while the velocity remained constant, gas clouds moving into and out of the line of sight with the same velocity is highly unlikely. Therefore, a change in ionizing flux is the likely cause for the trough depth variability. 
\\\\One concern \citet{sun17} had was the timescale needed for the outflow to respond was too long ($\sim$20~years). However, with the larger density determined in this work, the timescale is over a factor of 10 shorter when the same expression from \citet{sun17}, $t = (\alpha_i \sub{n}{e})^{-1}$, is used. Thus, the outflow response is quick enough for the optical depth changes seen in \citet{sun17}.
\subsection{Saturation and Contamination}
\begin{figure*}
\includegraphics[scale=1.0]{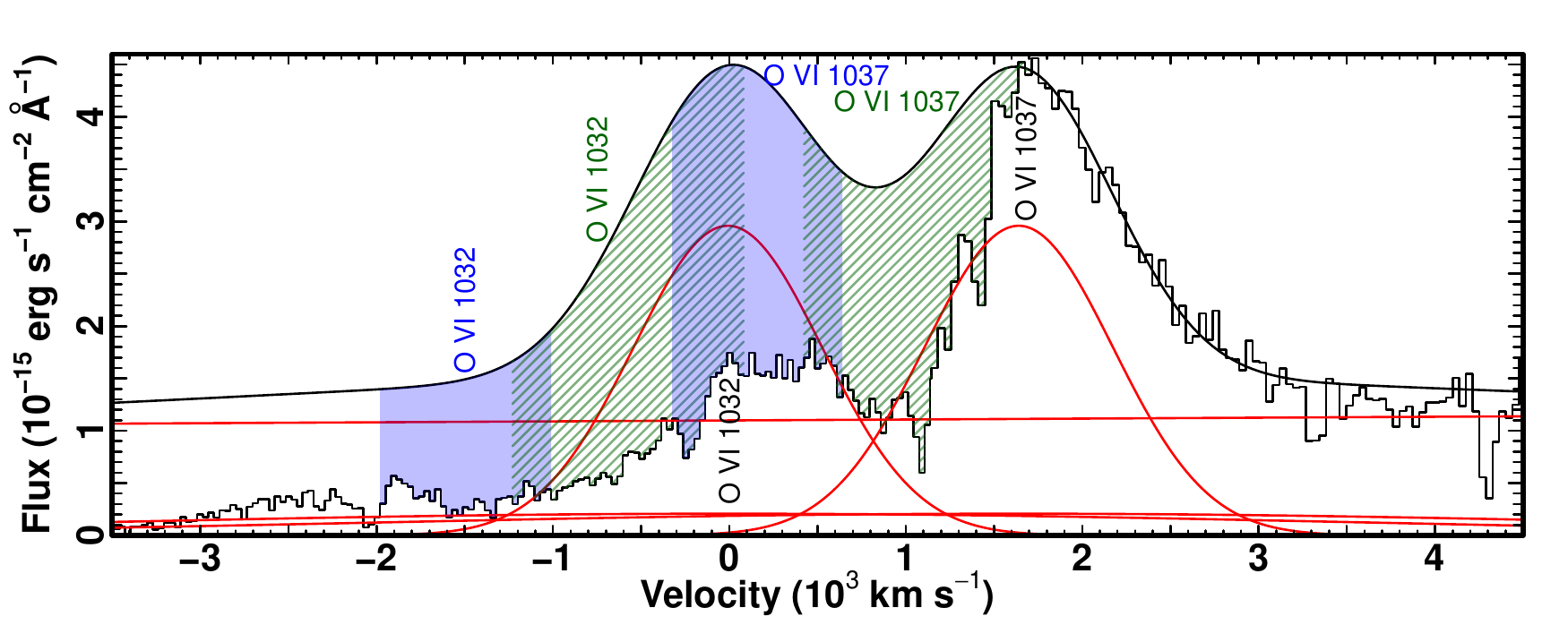}
\caption{Spectrum, unabsorbed emission model, and \sion{O}{vi} absorption troughs centered around the \sion{O}{vi} emission lines. In red, the unabsorbed emission model components of the power law with two broad and two intermediate-width Gaussian profiles.\label{fig:iwel}}
\end{figure*}

As stated in Section~\ref{sec:cd}, the column densities of \sion{Si}{ii} and \sion{O}{vi} were taken as measurements. Since the \sion{Si}{ii} column density ratio (1.76$^{+0.43}_{-0.30}$) is close to the maximum value physically allowed ($\sim$2), we believe the \sion{Si}{ii} column density is not significantly saturated, if at all. \citet{sun17} argued that the v1400 \sion{O}{vi} $\lambda$1032 trough is saturated. They explained the residual intensity of this trough as an intermediate-width emission line that is not covered by the v1400 outflow. However, the 2017 data cover the \sion{O}{vi} $\lambda$1037 emission for the first time, providing new constraints. Figure~\ref{fig:iwel} shows the spectrum, unabsorbed emission model, and \sion{O}{vi} absorption troughs centered around the \sion{O}{vi} emission lines as well as the decomposition of the unabsorbed emission model (a power law with two broad and two intermediate-width Gaussian profiles all in red). This model is minimalistic in that the peak of the \sion{O}{vi} $\lambda$1037 emission line is assumed to be the maximum and the \sion{O}{vi} $\lambda$1032 emission line is equal in strength. The \sion{O}{vi} $\lambda$1032 emission can be up to a factor of two higher (the oscillator strengths for these lines differ by this same factor) but cannot be weaker. 
\\\\Given the minimalistic model, the v1400 outflow must cover all components of the unabsorbed emission model, or else the residual intensity should trace the Gaussian profile for the \sion{O}{vi} $\lambda$1032 emission line. Therefore, the residual intensity is not from an uncovered emission region, as was previously thought by \citet{sun17}. Since the column density measurement for the v1400 \sion{O}{vi} $\lambda$1032 is less model dependent than the corresponding v1400 \sion{O}{vi} $\lambda$1037, we adopted its value and gave a conservative upper error of a factor of two to account for any saturation that may be present.
\\\\The contamination in the v1400 \sion{N}{iii} troughs by the v700 \sion{N}{iii} troughs affects the v1400 \sion{N}{iii} column density ratio (see Figure~\ref{fig:lines}). If the contamination is large enough, the v1400 \sion{N}{iii} column density ratio could fall sufficiently below 1 that there would be a discrepancy between the value for the electron number density for the v1400 outflow as determined by the \sion{N}{iii} and \sion{S}{iv} column density ratios. 
\\\\To estimate the amount of contamination, a constraint on the v700 \sion{N}{iii} column density ratio is needed. From \citet{sun17}, the hydrogen density, \sub{n}{H}, was estimated for the v700 outflow to be in the range of $10^9$--$10^{10}$ cm$^{-3}$. To corroborate this result, we calculated the \sion{S}{iv} column density ratio (Table~\ref{tab:ratios}) for the v700 outflow from the associated regions in Figure~\ref{fig:lines}. As can be seen in the lower half of Figure~\ref{fig:dens}, we derived a lower limit on the electron number density of log(\sub{n}{e}) $>$ 5.34~cm$^{-3}$ from the measured ratio and temperature range from the photoionization solutions in \citet{sun17} for the v700 outflow. Furthermore, we tested their best-fit model by measuring the column density of \sion{P}{v} $\lambda$1118 for the v700 outflow, a value of 17$^{+3}_{-3}\times$10$^{13}$ cm$^{-2}$, and found it to be consistent with the 16$\times$10$^{13}$ cm$^{-2}$ value predicted by the best-fit model. Therefore, the density range is sound, and the \sion{N}{iii} column density ratio for the v700 outflow must be in the range of 1--2, depending on the saturation. 
\\\\The most contamination that can occur is when the ratio is 1, maximizing the v700 \sion{N}{iii} resonance column density, which affects the value of the v1400 \sion{N}{iii} excited column density. To estimate the v700 \sion{N}{iii} column density for the resonance transition, we calculated the column density for the right half of the v700 \sion{N}{iii} excited transition (between 989 \AA~and 992 \AA~, rest-frame; Figure~\ref{fig:lines}), which suffers the least contamination from other troughs, and doubled it. This yielded a value of 90$\times 10^{13}$ cm$^{-2}$. Assuming this only affects the \sion{N}{iii} excited column density of the v1400 outflow (it would affect both) and ignoring saturation in the v1400 \sion{N}{iii} troughs, the v1400 \sion{N}{iii} column density ratio only decreases to 0.82$^{+0.17}_{-0.17}$, remaining consistent with a value of 1.   
\subsection{v700 Outflow ``Shading Effect" on the v1400 Outflow}
\label{sec:tp}
\begin{figure}
\includegraphics[scale=0.345]{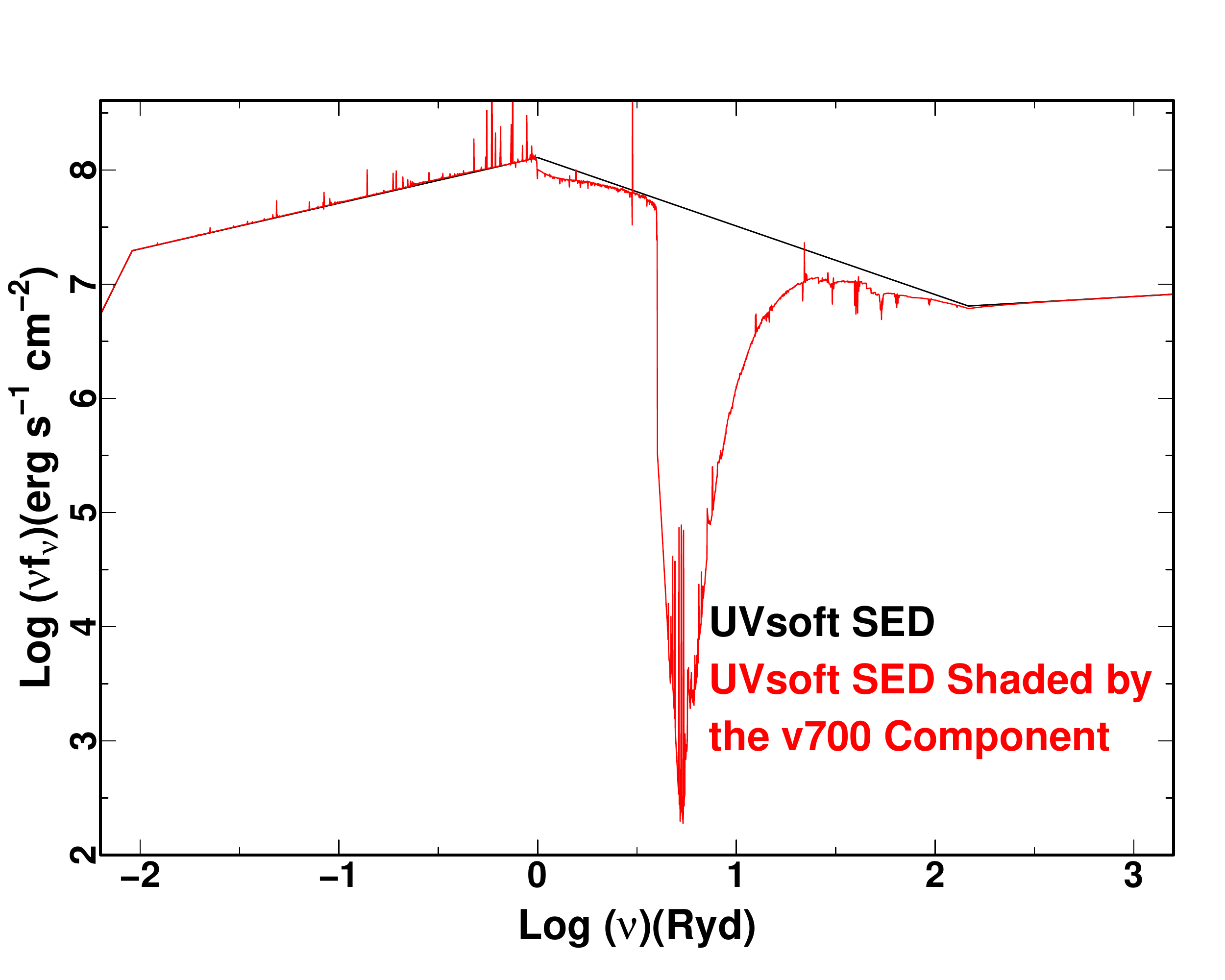}
\includegraphics[scale=0.305]{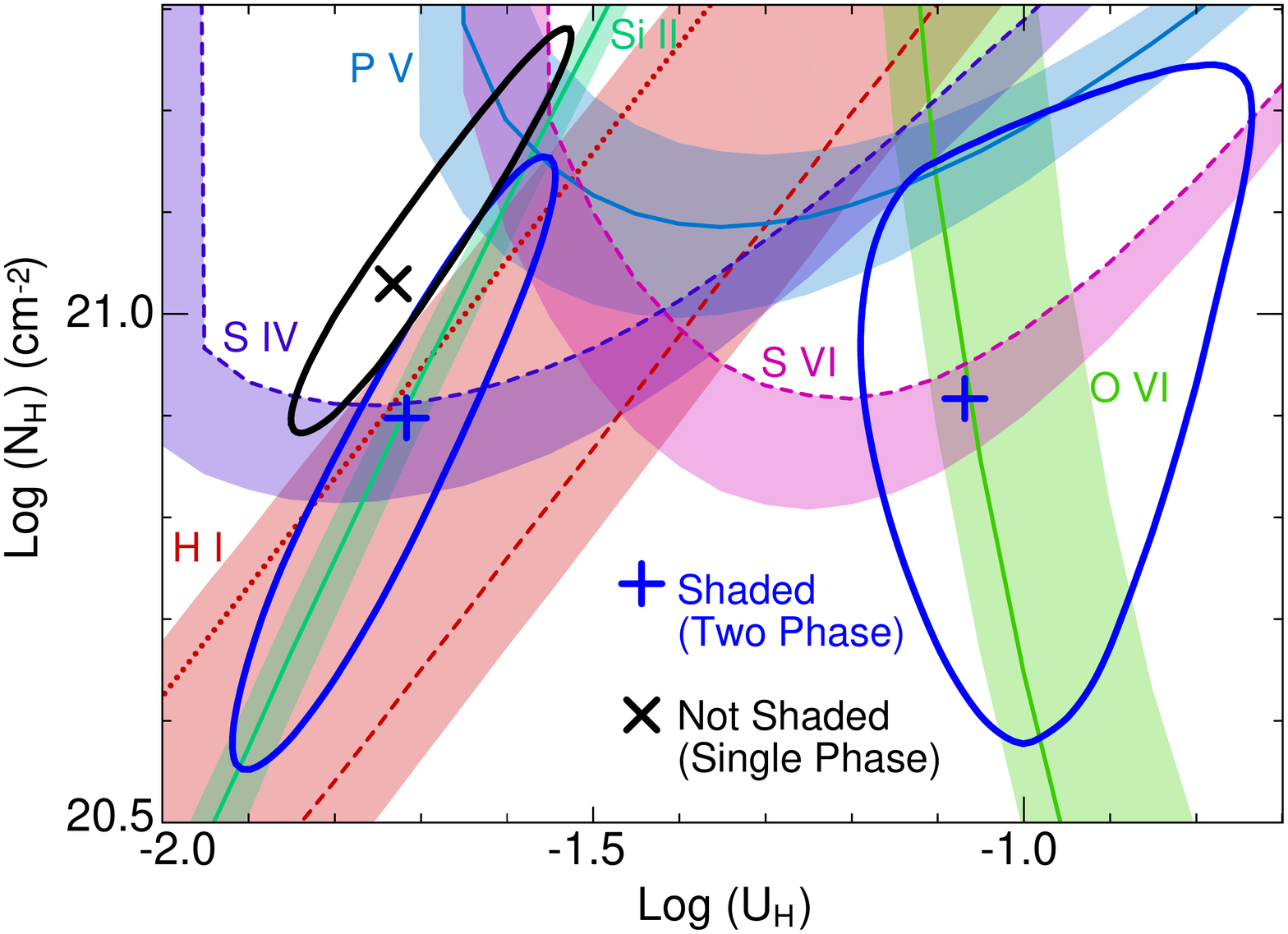}
\caption{\textit{Top:} comparison between the UV-soft SED and the transmitted SED after being shaded by the v700 outflow. \textit{Bottom:} two-phase photoionization solution for the v1400 outflow assuming the transmitted SED. The colored contours, shaded bands, symbols, and ellipses have the same meanings as in Figure~\ref{fig:sol}.\label{fig:tp}}
\end{figure}

To test the effects the v700 ``shading" has on the v1400 outflow, we first used the best-fit parameters of \citet{sun17} for the v700 outflow and generated a Cloudy model, assuming the UV-soft SED. From this model, we used the transmitted SED as the input SED to generate a new grid of models to find a new photoionization solution for the v1400 outflow. The original UV-soft SED and shaded SED can be seen in the top panel of Figure~\ref{fig:tp}. The major difference is the decrease in photons between 4 and 100 Rydbergs. This decrease in photons affects the ionization parameter required to match the column densities of high-ionization ions like \sion{O}{vi} as seen in the bottom panel of Figure~\ref{fig:tp}. The \sion{O}{vi} contour has shifted by $\sim$0.7 dex in ionization parameter. This shift necessitates a two-phase photoionization solution to satisfy the observed ionic column densities: one component with log(\sub{N}{H})~$=20.90^{+0.26}_{-0.35}$~cm$^{-2}$ and log(U$_H$)~$=-1.72^{+0.17}_{-0.20}$; and a second component with log(\sub{N}{H})~$=20.92^{+0.33}_{-0.34}$~cm$^{-2}$ and log(U$_H$)~$=-1.07^{+0.33}_{-0.12}$. A single-phase photoionization solution cannot satisfy the ionic column densities of \sion{Si}{ii} and \sion{O}{vi} simultaneously without overpredicting the \sion{P}{v} column density by an order of magnitude. 
\\\\The lower-ionization parameter solution contains nearly all of the \sion{S}{iv} column density, and therefore corresponds to the same density as in the unshaded solution. This allows for the distance to be determined. Using the same normalization as in Section~\ref{v1400opde} and the shaded SED, we calculate $\sub{Q}{H}=1.72\times10^{56}$~s$^{-1}$ and R=420$^{+130}_{-80}$~pc. This yields a mass flux and kinetic luminosity of $\dot{M}$ = 11.3$^{+8.1}_{-5.5}$ and log($\dot{\sub{E}{K}}$) = 42.9$^{+0.23}_{-0.29}$ erg s$^{-1}$, respectively. This distance, mass flux, and kinetic luminosity are all consistent with the values determined for the unshaded photoionization solution. The main difference is that the shaded solution requires a two-phase ionization outflow with $\Delta U \approxeq 0.7$.

\section{Summary and Conclusions}
\label{sec:sc}
In this paper, we presented new \textit{HST}/COS spectra for the quasar outflows seen in LBQS 1206+1052. We identified absorption troughs in this spectra from ions \sion{H}{i}, \sion{C}{iii}, \sion{N}{iii}, \sion{N}{v}, \sion{O}{iii}, \sion{O}{vi}, \sion{S}{iii}, \sion{S}{iv}, \sion{S}{vi}, \sion{Si}{ii}, \sion{Si}{iii}, and \sion{P}{v}. Then, we calculated ionic column densities for the v1400 outflow from these data as well as reprocessed, archival \textit{HST}/COS data (Table~\ref{tab:col}). From these ionic column densities and a grid of photoionization models, we determined the best-fit ionization parameter and hydrogen column density for each dataset: log(\sub{U}{H})$=-1.75^{+0.20}_{-0.12}$ and log(\sub{N}{H})$=20.96^{+0.21}_{-0.16}$~cm$^{-2}$ for the 2010 data and  log(\sub{U}{H})$=-1.73^{+0.21}_{-0.12}$ and log(\sub{N}{H})$=21.03^{+0.25}_{-0.15}$~cm$^{-2}$ for the 2017 data. 
\\\\For the ions \sion{S}{iv}, \sion{Si}{ii}, \sion{S}{iii}, and \sion{N}{iii}, we detected troughs from both resonance and excited states, the ratios of which are density sensitive. These ratios were compared to theoretical models of the ratio versus electron number density. The \sion{S}{iv} column density ratio provided the adopted electron number density, log(\sub{n}{e}) = 4.23$^{+0.09}_{-0.09}$~cm$^{-3}$. The column density ratios from \sion{N}{iii} and \sion{S}{iii} were consistent with a value of 1, indicative of saturation, and therefore only gave lower limits to \sub{n}{e}. The electron number density determined from the \sion{Si}{ii} column density ratio was also taken as a lower limit since the upper limit of the ratio was consistent with \sub{n}{e} being much larger than the critical density. From the adopted electron number density, the distance of the v1400 outflow from the central source was calculated with Equation (\ref{eq:R}), yielding R = 500$^{+100}_{-110}$ pc. Using this distance and Equations (\ref{eq:M})~\&~(\ref{eq:E}), the mass flux and kinetic luminosity were also determined to be $\dot{M}$ = 8.9$^{+7.2}_{-3.1}$ $M_{\astrosun}$yr$^{-1}$ and log($\dot{\sub{E}{K}}$) = 42.8$^{+0.26}_{-0.19}$ erg s$^{-1}$, respectively.
\\\\The following conclusions emerge from this work:
\begin{enumerate}
\item The new observations, with the wider wavelength coverage, were paramount in determining that the correct electron number density for the v1400 outflow in LBQS 1206+1052 is over an order of magnitude larger than previously thought. This coupled with accounting for dust extinction decreased the distance by nearly a factor of two.
\item Using a shaded SED leads to a two-phase outflow for the v1400 component but has little effect on $R$, $\dot{M}$, and $\dot\sub{E}{K}$.
\item For the v700 outflow, using the \sion{S}{iv} ratio, we set a lower limit on the electron number density that is consistent with previous results.
\end{enumerate}



\acknowledgments

T.M. and N.A. acknowledge support from NASA  grants \textit{HST} GO 14777, 14242, 14054, and 14176. This support is provided by NASA through a grant from the Space Telescope Science Institute, which is operated by the Association of Universities for Research in Astronomy, Incorporated, under NASA contract NAS5-26555. T.M. and N.A. also acknowledge support from NASA ADAP 48020 and NSF grant AST 1413319. All authors are grateful to their home institutions for travel support, if provided, and to the anonymous referee whose careful review improved the quality of this paper. G.L. acknowledges the grant from the National Key R\&D Program of China (2016YFA0400702), the National Natural Science Foundation of China (No. 11673020 and No. 11421303), and the National Thousand Young Talents Program of China. CHIANTI is a collaborative project involving George Mason University, the University of Michigan (USA), and the University of Cambridge (UK).




\begin{thebibliography}{}

\bibitem[Aldcroft et al.(1994)]{ald94} Aldcroft, T.~L., Bechtold, J., \& Elvis, M.\ 1994, \apjs, 93, 1 
\bibitem[Angl{\'e}s-Alc{\'a}zar et al.(2017)]{ang17} Angl{\'e}s-Alc{\'a}zar, D., Dav{\'e}, R., Faucher-Gigu{\`e}re, C.-A., {\"O}zel, F., \& Hopkins, P.~F.\ 2017, \mnras, 464, 2840 
\bibitem[Aoki et al.(2011)]{aok11} Aoki, K., Oyabu, S., Dunn, J.~P., et al.\ 2011, \pasj, 63, 457 
\bibitem[Arav et al.(2008)]{ara08} Arav, N., Moe, M., Costantini, E., et al.\ 2008, \apj, 681, 954-964 
\bibitem[Arav et al.(2012)]{ara12} Arav, N., Edmonds, D., Borguet, B., et al.\ 2012, \aap, 544, A33 
\bibitem[Arav et al.(2013)]{ara13} Arav, N., Borguet, B., Chamberlain, C., Edmonds, D., \& Danforth, C.\ 2013, \mnras, 436, 3286
\bibitem[Arav et al.(2015)]{ara15} Arav, N., Chamberlain, C., Kriss, G.~A., et al.\ 2015, \aap, 577, A37
\bibitem[Arav et al.(2018)]{ara18} Arav, N., Liu, G., Xu, X., et al.\ 2018, \apj, 857, 60 
\bibitem[Barai et al.(2011)]{bar11} Barai, P., Proga, D., \& Nagamine, K.\ 2011, \mnras, 418, 591 
\bibitem[Bautista et al.(2010)]{bau10} Bautista, M.~A., Dunn, J.~P., Arav, N., et al.\ 2010, \apj, 713, 25
\bibitem[Bennett et al.(2014)]{ben14} Bennett, C.~L., Larson, D., Weiland, J.~L., \& Hinshaw, G.\ 2014, \apj, 794, 135  
\bibitem[Blandford \& Begelman(2004)]{bla04} Blandford, R.~D., \& Begelman, M.~C.\ 2004, \mnras, 349, 68 
\bibitem[Booth \& Schaye(2009)]{boo09} Booth, C.~M., \& Schaye, J.\ 2009, American Institute of Physics Conference Series, 1201, 21 
\bibitem[Borguet et al.(2012)]{bor12} Borguet, B.~C.~J., Edmonds, D., Arav, N., Dunn, J., \& Kriss, G.~A.\ 2012, \apj, 751, 107 
\bibitem[Borguet et al.(2012b)]{bor12b} Borguet, B.~C.~J., Edmonds, D., Arav, N., Benn, C., \& Chamberlain, C.\ 2012b, \apj, 758, 69
\bibitem[Borguet et al.(2013)]{bor13} Borguet, B.~C.~J., Arav, N., Edmonds, D., Chamberlain, C., \& Benn, C.\ 2013, \apj, 762, 49 
\bibitem[Capellupo et al.(2013)]{cap13} Capellupo, D.~M., Hamann, F., Shields, J.~C., Halpern, J.~P., \& Barlow, T.~A.\ 2013, \mnras, 429, 1872
\bibitem[Chamberlain et al.(2015)]{cha15a} Chamberlain, C., Arav, N., \& Benn, C.\ 2015, \mnras, 450, 1085 
\bibitem[Chamberlain \& Arav(2015)]{cha15b} Chamberlain, C., \& Arav, N.\ 2015, \mnras, 454, 675 
\bibitem[Choi et al.(2017)]{cho17} Choi, E., Ostriker, J.~P., Naab, T., et al.\ 2017, \apj, 844, 31 
\bibitem[Ciotti et al.(2009)]{cio09} Ciotti, L., Ostriker, J.~P., \& Proga, D.\ 2009, \apj, 699, 89 
\bibitem[Dai et al.(2008)]{dai08} Dai, X., Shankar, F., \& Sivakoff, G.~R.\ 2008, \apj, 672, 108-114 
\bibitem[de Kool et al.(2001)]{dek01} de Kool, M., Arav, N., Becker, R.~H., et al.\ 2001, \apj, 548, 609 
\bibitem[de Kool et al.(2002)]{dek02} de Kool, M., Becker, R.~H., Arav, N., Gregg, M.~D., \& White, R.~L.\ 2002, \apj, 570, 514 
\bibitem[Dere et al.(1997)]{der97} Dere, K.~P., Landi, E., Mason, H.~E., Monsignori Fossi, B.~C., \& Young, P.~R.\ 1997, \aaps, 125, 149
\bibitem[Dubois et al.(2014)]{dub14} Dubois, Y., Volonteri, M., \& Silk, J.\ 2014, \mnras, 440, 1590 
\bibitem[Dunn et al.(2010)]{dun10} Dunn, J.~P., Bautista, M., Arav, N., et al.\ 2010, \apj, 709, 611 
\bibitem[Edmonds et al.(2011)]{edm11} Edmonds, D., Borguet, B., Arav, N., et al.\ 2011, \apj, 739, 7 
\bibitem[Ely et al.(2011)]{ely11} Ely, J., Massa, D., Ake, T., et al.\ 2011, COS FUV Gridwire Flat Field Template, Instrument Science Report COS 2011-03, Space Telescope Science Institute (Baltimore, MD: STScI), http://www.stsci.edu/hst/cos/documents/isrs/
ISR2011\_03.pdf
\bibitem[Ferland et al.(2017)]{fer17} Ferland, G.~J., Chatzikos, M., Guzm{\'a}n, F., et al.\ 2017, \rmxaa, 53, 385
\bibitem[Filiz Ak et al.(2013)]{fil13} Filiz Ak, N., Brandt, W.~N., Hall, P.~B., et al.\ 2013, \apj, 777, 168 
\bibitem[Finn et al.(2014)]{fin14} Finn, C.~W., Morris, S.~L., Crighton, N.~H.~M., et al.\ 2014, \mnras, 440, 3317 
\bibitem[Gabel et al.(2005)]{gab05} Gabel, J.~R., Kraemer, S.~B., Crenshaw, D.~M., et al.\ 2005, \apj, 631, 741 
\bibitem[Gibson et al.(2008)]{gib08} Gibson, R.~R., Brandt, W.~N., Schneider, D.~P., \& Gallagher, S.~C.\ 2008, \apj, 675, 985-1001
\bibitem[Gibson et al.(2009)]{gib09} Gibson, R.~R., Jiang, L., Brandt, W.~N., et al.\ 2009, \apj, 692, 758
\bibitem[Grevesse et al.(2010)]{gre10} Grevesse, N., Asplund, M., Sauval, A.~J., \& Scott, P.\ 2010, \apss, 328, 179
\bibitem[Grier et al.(2015)]{gri15} Grier, C.~J., Hall, P.~B., Brandt, W.~N., et al.\ 2015, \apj, 806, 111 
\bibitem[Grier et al.(2016)]{gri16} Grier, C.~J., Brandt, W.~N., Hall, P.~B., et al.\ 2016, \apj, 824, 130 
\bibitem[Hall et al.(2011)]{hal11} Hall, P.~B., Anosov, K., White, R.~L., et al.\ 2011, \mnras, 411, 2653 
\bibitem[Hamann et al.(1997)]{ham97} Hamann, F., Barlow, T., Cohen, R.~D., Junkkarinen, V., \& Burbidge, E.~M.\ 1997, Mass Ejection from Active Galactic Nuclei, Astronomical Society of the Pacific Conference Series, 128, 19 
\bibitem[Hamann et al.(2001)]{ham01} Hamann, F.~W., Barlow, T.~A., Chaffee, F.~C., Foltz, C.~B., \& Weymann, R.~J.\ 2001, \apj, 550, 142
\bibitem[Hamann \& Sabra(2004)]{ham04} Hamann, F., \& Sabra, B.\ 2004, AGN Physics with the Sloan Digital Sky Survey, Astronomical Society of the Pacific Conference Series, 311, 203 
\bibitem[Hamann et al.(2008)]{ham08} Hamann, F., Kaplan, K.~F., Rodr{\'{\i}}guez Hidalgo, P., Prochaska, J.~X., \& Herbert-Fort, S.\ 2008, \mnras, 391, L39 
\bibitem[He et al.(2017)]{he17} He, Z., Wang, T., Zhou, H., et al.\ 2017, \apjs, 229, 22 
\bibitem[Hewett et al.(1995)]{hew95} Hewett, P.~C., Foltz, C.~B., \& Chaffee, F.~H.\ 1995, \aj, 109, 1498 
\bibitem[Hewett \& Foltz(2003)]{hew03} Hewett, P.~C., \& Foltz, C.~B.\ 2003, \aj, 125, 1784 
\bibitem[Hopkins et al.(2009)]{hop09} Hopkins, P.~F., Murray, N., \& Thompson, T.~A.\ 2009, \mnras, 398, 303 
\bibitem[Ji et al.(2012)]{ji12} Ji, T., Wang, T.-G., Zhou, H.-Y., \& Wang, H.-Y.\ 2012, Research in Astronomy and Astrophysics, 12, 369
\bibitem[Khalatyan et al.(2008)]{kha08} Khalatyan, A., Cattaneo, A., Schramm, M., et al.\ 2008, \mnras, 387, 13  
\bibitem[Knigge et al.(2008)]{kni08} Knigge, C., Scaringi, S., Goad, M.~R., \& Cottis, C.~E.\ 2008, \mnras, 386, 1426 
\bibitem[Korista et al.(2008)]{kor08} Korista, K.~T., Bautista, M.~A., Arav, N., et al.\ 2008, \apj, 688, 108-115
\bibitem[Landi et al.(2013)]{lan13} Landi, E., Young, P.~R., Dere, K.~P., Del Zanna, G., \& Mason, H.~E.\ 2013, \apj, 763, 86 
\bibitem[Li et al.(2015)]{li15} Li, Z., Zhou, H., Hao, L., et al.\ 2015, \apj, 812, 99
\bibitem[Lucy et al.(2014)]{luc14} Lucy, A.~B., Leighly, K.~M., Terndrup, D.~M., Dietrich, M., \& Gallagher, S.~C.\ 2014, \apj, 783, 58
\bibitem[Lundgren et al.(2007)]{lun07} Lundgren, B.~F., Wilhite, B.~C., Brunner, R.~J., et al.\ 2007, \apj, 656, 73
\bibitem[Mathews \& Ferland(1987)]{mat87} Mathews, W.~G., \& Ferland, G.~J.\ 1987, \apj, 323, 456  
\bibitem[McGraw et al.(2017)]{mcg17} McGraw, S.~M., Brandt, W.~N., Grier, C.~J., et al.\ 2017, \mnras, 469, 3163 
\bibitem[McGraw et al.(2018)]{mcg18} McGraw, S.~M., Shields, J.~C., Hamann, F.~W., Capellupo, D.~M., \& Herbst, H.\ 2018, \mnras, 475, 585 
\bibitem[Moe et al.(2009)]{moe09} Moe, M., Arav, N., Bautista, M.~A., \& Korista, K.~T.\ 2009, \apj, 706, 525
\bibitem[Murray et al.(1995)]{mur95} Murray, N., Chiang, J., Grossman, S.~A., \& Voit, G.~M.\ 1995, \apj, 451, 498 
\bibitem[Peirani et al.(2017)]{pei17} Peirani, S., Dubois, Y., Volonteri, M., et al.\ 2017, \mnras, 472, 2153   
\bibitem[Proga et al.(2000)]{pro00} Proga, D., Stone, J.~M., \& Kallman, T.~R.\ 2000, \apj, 543, 686 
\bibitem[Proga \& Kallman(2004)]{pro04} Proga, D., \& Kallman, T.~R.\ 2004, \apj, 616, 688 
\bibitem[Reichard et al.(2003)]{rei03} Reichard, T.~A., Richards, G.~T., Hall, P.~B., et al.\ 2003, \aj, 126, 2594 
\bibitem[Rosas-Guevara et al.(2015)]{ros15} Rosas-Guevara, Y.~M., Bower, R.~G., Schaye, J., et al.\ 2015, \mnras, 454, 1038 
\bibitem[Savage \& Sembach(1991)]{sav91} Savage, B.~D., \& Sembach, K.~R.\ 1991, \apj, 379, 245 
\bibitem[Scannapieco \& Oh(2004)]{sca04} Scannapieco, E., \& Oh, S.~P.\ 2004, \apj, 608, 62 
\bibitem[Schaye et al.(2015)]{sch15} Schaye, J., Crain, R.~A., Bower, R.~G., et al.\ 2015, \mnras, 446, 521 
\bibitem[Shang et al.(2005)]{sha05} Shang, Z., Brotherton, M.~S., Green, R.~F., et al.\ 2005, \apj, 619, 41.
\bibitem[Shang et al.(2011)]{sha11} Shang, Z., Brotherton, M.~S., Wills, B.~J., et al.\ 2011, The Astrophysical Journal Supplement Series, 196, 2.
\bibitem[Silk \& Rees(1998)]{sil98} Silk, J., \& Rees, M.~J.\ 1998, \aap, 331, L1 
\bibitem[STScI Newsletter(2016)]{sts16} Space Telescope Science Institute Newsletter\ 2016, Hubble Space Telescope Updated COS/FUV Wavelength Dispersion Solution Reference File (DISPTAB) Released, Space Telescope Science Institute (Baltimore, MD: STScI), http://www.stsci.edu/hst/cos/documents/
newsletters/cos\_newsletters/full\_stories/
2016\_05/new\_disptab
\bibitem[Sun et al.(2017)]{sun17} Sun, L., Zhou, H., Ji, T., et al.\ 2017, \apj, 838, 88
\bibitem[Taylor \& Kobayashi(2015)]{tay15} Taylor, P., \& Kobayashi, C.\ 2015, \mnras, 452, L59 
\bibitem[Tornatore et al.(2010)]{tor10} Tornatore, L., Borgani, S., Viel, M., \& Springel, V.\ 2010, \mnras, 402, 1911
\bibitem[Trump et al.(2006)]{tru06} Trump, J.~R., Hall, P.~B., Reichard, T.~A., et al.\ 2006, \apjs, 165, 1 
\bibitem[Vestergaard(2003)]{ves03} Vestergaard, M.\ 2003, \apj, 599, 116   
\bibitem[Vilkoviskij \& Irwin(2001)]{vil01} Vilkoviskij, E.~Y., \& Irwin, M.~J.\ 2001, \mnras, 321, 4
\bibitem[Vivek et al.(2012)]{viv12} Vivek, M., Srianand, R., Mahabal, A., \& Kuriakose, V.~C.\ 2012, \mnras, 421, L107  
\bibitem[Volonteri et al.(2016)]{vol16} Volonteri, M., Dubois, Y., Pichon, C., \& Devriendt, J.\ 2016, \mnras, 460, 2979 
\bibitem[Wang et al.(2015)]{wan15} Wang, T., Yang, C., Wang, H., \& Ferland, G.\ 2015, \apj, 814, 150 
\bibitem[Weymann et al.(1991)]{wey91} Weymann, R.~J., Morris, S.~L., Foltz, C.~B., \& Hewett, P.~C.\ 1991, \apj, 373, 23
\bibitem[Wildy et al.(2015)]{wil15} Wildy, C., Goad, M.~R., \& Allen, J.~T.\ 2015, \mnras, 448, 2397 
\bibitem[Xu et al.(2018)]{xu18} Xu, X., Arav, N., Miller, T., \& Benn, C.\ 2018, \apj, 858, 39 

\end{thebibliography}
\end{document}